\documentclass[fleqn,usenatbib,useAMS,twocolumn]{mnras}
\usepackage{graphicx}	
\usepackage{amsmath}	
\usepackage{amssymb}	
\usepackage{amsfonts}
\usepackage{amsbsy}
\usepackage{color}
\usepackage{rotating}
\usepackage{multicol}        
\usepackage{dcolumn}
\usepackage{bm}		
\usepackage{lscape}	
\usepackage[T1]{fontenc}
\usepackage[english]{babel}
\usepackage{hyperref}
\usepackage{ae,aecompl}
\usepackage{cleveref}
\usepackage[rightcaption]{sidecap}
\usepackage{subfigure}
\usepackage{longtable}

\title[Dark energy models with the $E_{\rm p,i}$ -- $E_{\rm iso}$ correlation ]{Prospects of high redshift constraints on
dark energy models with the $E_{\rm p,i}$ -- $E_{\rm iso}$ correlation in long Gamma Ray Bursts}

\author[M. Demianski et al.]{
M.Demianski,$^{1,2}$
E.Piedipalumbo,$^{3,4}$\thanks{E-mail: ester@na.infn.it}
D.Sawant$^{5,6}$
and L.Amati $^{7}$
\\
$^{1}$Institute for Theoretical Physics, University of Warsaw, Pasteura 5, 02-093 Warsaw, Poland\\
$^{2}$Department of Astronomy, Williams College, Williamstown, MA 01267, USA\\
$^{3}$Dipartimento di Fisica, Universit\`{a} degli Studi di Napoli Federico II, Compl. Univ. Monte S. Angelo, 80126 Naples, Italy\\
$^{4}$I.N.F.N., Sez. di Napoli, Compl. Univ. Monte S. Angelo, Edificio 6, via Cinthia, 80126 - Napoli, Italy\\
$^{5}$Dipartimento di Fisica e Scienze della Terra, Universit\`{a} degli Studi di Ferrara\\
$^{6}$ IIT, Mombay, India\\
$^{7}$INAF-IASF, Sezione di Bologna, via Gobetti 101, 40129 Bologna, Italy}

\date{Accepted XXX. Received YYY; in original form ZZZ}

\pubyear{2021}

\begin{document}
\maketitle

\begin{abstract}
So far large and different data sets revealed the accelerated expansion rate of the
Universe, which is usually explained in terms of dark energy. The nature of dark energy is not yet
known, and several models have been introduced: a non zero cosmological constant, a potential
energy of some scalar field, effects related to the non homogeneous distribution of matter, or
effects due to alternative theories of gravity. Recently, a tension with the flat $\Lambda$CDM model has been discovered using a high-redshift Hubble diagram of supernovae, quasars, and gamma-ray bursts. Here we use the Union2 type Ia supernovae  (SNIa) and Gamma Ray Bursts (GRB) Hubble diagram, and a set of direct measurements of the Hubble parameter to explore different dark energy models. We use the Chevallier-Polarski- Linder
(CPL) parametrization of the dark energy equation of state (EOS), a minimally coupled  quintessence scalar field, and, finally, we
consider models  with dark energy at early times (EDE). We perform a statistical analysis based on the Markov
chain Monte Carlo (MCMC) method, and explore the probability distributions
of the cosmological parameters for each of the competing models. We apply the Akaike Information
Criterion (AIC) to compare these models: our analysis indicates that an evolving dark energy,  described by a scalar field  with exponential potential seems to be favoured by observational data.
\end{abstract}




\section{Introduction}

Starting at the end of the 1990s, observations of high-redshift supernovae of type Ia (SNIa) revealed  the
accelerated expansion of the Universe \citep[][]{ perl98, per+al99, Riess, Riess07, SNLS,Union2}. This unexpected result has been confirmed by statistical analysis of observations of
small-scale temperature anisotropies of
the Cosmic Microwave Background Radiation (CMBR) \citep[][]{WMAP13, PlanckXIII}. The observed accelerated expansion is
usually related to a non zero cosmological constant or to existence of so called dark energy, a cosmic medium with positive energy density but sufficiently large negative pressure, which now provides
about $70\% $ of the matter
energy in the Universe. The nature of dark energy is, however, not known. The models of dark energy proposed so far
range from a non-zero cosmological constant \citep[][]{Peebles84, carroll01}, to a potential energy of some not
yet discovered scalar field  \citep[][]{SF, alma}, to effects connected with inhomogeneous distribution
of matter and averaging procedures \citep[][]{clark}, or modifications of the Einstein General Theory of Relativity \cite[][]{DeFelice, capozziello11, clifton, capozziello19}. In the last  cases, in general, the effective EOS of dark energy is
not constant, but depends on redshift $z$.   Therefore populating the Hubble diagram up to high redshifts remains a primary task to test the consistency of the $\Lambda$CDM model  \citep[see for instance][for discussions about the  $\Lambda$CDM tension ]{Lusso19,risaliti1,Lusso20,Lusso20b} and, therefore, to shine new light on the nature of dark energy. So far dark energy models are poorly tested in the redshift interval between the farthest observed Type Ia supernovae and that of the Cosmic Microwave Background. In our high redshift investigation we consider the Union2 SNIa data set, and the long gamma-ray
burst (GRB) Hubble diagram, constructed by calibrating the correlation between the peak
photon energy, $E_{\mathrm{p, i}}$, and the isotropic equivalent radiated energy, $ E_{\mathrm{iso}}$
\citep[]{{MGRB1}, {MGRB2b}}.
Here we consider an {\it extended } $E_{\rm p,i} $- $E_{\rm iso} $ correlation, to take into account  possible redshift evolution effects, modeled  through power law terms. It turns out that at least a part of this evolution can be caused by gravitational lensing effects \citep[see, for instance,][]{Shirokov20}.  We consider also a sample of 28 direct measurements of the Hubble parameter, compiled by \citep[][]{farooqb}. These measurements are performed through the differential age technique, first suggested by  \citep[][]{Jimenez}, which uses red passively evolving galaxies as cosmic chronometers.
Here we probe the dynamical evolution of dark energy, by considering some proposed so far competing models of dark energy:
\begin{itemize}
\item[i)] an empirically parametrized EOS of dark energy, usually using two or more free parameters.
Among all the proposed parametrization forms of the dark energy  EOS, we consider the CPL
\citep[][]{cpl1, cpl2}, which is now widely used\,,
\item[ii)] a  quintessence dark energy: a model where a self interacting scalar field plays the role of dark
energy and drives the acceleration \citep[][]{Peebles88a, Peebles88b, Tsujikawa13}\,,
\item[iii)] early dark energy: models where a non negligible fraction of dark energy exists already at early stages of evolution of the Universe \citep[][]{Khoraminazad20}\,.
\end{itemize}


Our statistical analysis is based on the Monte Carlo Markov Chain (MCMC) simulations to simultaneously compute the full
probability density functions (PDFs) of all the parameters of interest. The structure of the paper is as follows. In
Sect. 2 we describe the different models of dark energy considered in our analysis.
 In Sect. 3 we describe the observational data sets that are used in
our analysis. In Sect. 4 we describe some details of our statistical analysis and
present results. In
Sect. 5 we present constrains on dark energy models that could be derived from future GRB Hubble diagram samples.
 General discussion of our results and conclusions are presented in
Sect. 6.

\section{Different models of dark energy}
Although seemingly consistent with the current standard model where
the cosmic acceleration is due to the Einstein's cosmological
constant, $\Lambda$, the precision of current data is not
sufficient to rule out  an evolving dark energy term. If then
the cosmological constant is not responsible for the observed accelerated
expansion of the Universe, we are considering some of the proposed models of a
dynamical field that is generating an effective negative
pressure. Moreover this could also indicate that the
cosmological Copernican principle cannot be applied at certain scales, and
that radial inhomogeneities could mimic the accelerated expansion.
Within the Friedman-Lemaitre-Robertson-Walker (FLRW) paradigm,
all possibilities can be characterized, as far as the background
dynamics is concerned, by the dark energy EOS, $w(z)$. The main
task of observational cosmology  is to search for evidence for
$w(z) \neq -1$.  This is usually done in terms of an appropriate
parameterization of the EOS.
\subsection{Parametrization of the dark energy EOS}
Within the Friedman equations dark energy appears through its
effective energy density, $\rho_{de}$, and pressure, $p_{de}$:

\begin{equation}
H^2 = \frac{8 \pi G}{3} (\rho_m + \rho_{de}) \,,
\label{eq: fried1}
\end{equation}

\begin{equation}
\frac{\ddot{a}}{a} = - \frac{4 \pi G}{3} \ \left(\rho_m + \rho_{de} + 3 p_{de}\right) \ ,
\label{eq: fried2}
\end{equation}

where $a$ is the scale factor,  $H = \dot{a}/a$ the Hubble
parameter, and  $\mathbf{\rho}_{m}$  is the dark matter energy density. Here
and throughout this paper the dot denotes the derivative with respect to
the cosmic time, and we have assumed a spatially flat Universe
in agreement with what is inferred from the CMBR anisotropy spectrum
\citep[][]{PlanckXIII}. The continuity equation for any cosmological
fluid is \,:
\begin{equation}
\frac{\dot{\rho_i}}{\rho_i} = - 3 H \left(1 +
\frac{p_i}{\rho_i}\right) = - 3 H  \left[1 + w_i(t)\right]\,,
\label{eq: continuity}
\end{equation}
where the energy density  $\rho_i$, the pressure $p_i$, and the EOS of each component is defined by
$\displaystyle{w_i = \frac{p_i}{\rho_i}}$. For ordinary nonrelativistic matter $w=0$, and the cosmological constant
can be treated as a medium with
$w=-1$. Let us recall that $\rho_m = D a^{-3}$, where the parameter $D \equiv \rho_{m0}{a_0}^3$ is determined by the
current values of $\rho_m$ and $a$. If we explicitly allow the possibility that the dark energy evolves, the importance
of its equation of state is significant and it determines the Hubble function $H(z)$, and any derivation of it as
needed to obtain the observable quantities. Actually it turns out that:
\begin{eqnarray}\label{heos}
  H(z,{\mathrm \theta}) &=& H_0 \sqrt{(1-\Omega_m) g(z, {\mathrm \theta})+\Omega_m (z+1)^3}\,,\nonumber\\
\end{eqnarray}
 where
 \begin{equation}
 g(z, {\mathrm \theta})=\frac{\rho_{de}(z,{\mathrm \theta} )}{\rho_{de}(0)}=e^{3 \int_0^z \frac{w(x,{\mathrm \theta})+1}{x+1} \, dx}\,,\nonumber\\
 \end{equation}
  $w(z,{\mathrm \theta})$ is any dynamical form of the dark energy EOS, and ${\mathrm \theta}=(\theta_1, \theta_1..,\theta_n)$ are the EOS  parameters.

It is worth noting that in Eq. (\ref{heos}) we are neglecting the radiation term $\Omega_r (1+z)^4$, but in this way  we introduce errors much smaller than the observational uncertainties of all our datasets.  Of course in the case of the early dark energy, when we have to investigate a dark energy component just in early times, we must include the radiation term.

Using the Hubble function we define the luminosity distance $d_L$ as
\begin{eqnarray}
d_L(z,{\mathrm \theta}) &=&\frac{c}{H_0} (1+z)\int^{z}_{0}{dy \over H(y,{\mathrm \theta})}\, \label{lumdgen}\\
&=&\frac{c}{H_0} (1+z)\int_0^z \frac{dy}{  \sqrt{(1-\Omega_m) g(y,{\mathrm \theta})+\Omega_m (y+1)^3}}\,.\label{lumd} \nonumber \\
\end{eqnarray}
Using the luminosity distance, we can evaluate the distance
modulus, from its standard definition (in Mpc):
\begin{equation}
\mu(z) = 25 + 5 \log{d_L(z,{\mathrm{\theta}})}.
\label{eq:defmu}
\end{equation}

In this work we consider the so-called  CPL parametrization of
the dark energy EOS \citep[][]{cpl1, cpl2} given by
\begin{equation}
w(z) =w_0 + w_{1} z (1 + z)^{-1} \,,
\label{cpleos}
\end{equation}
 where $w_0$ and $w_{1}$ are real numbers that represent the EOS present value and its overall time evolution,
respectively.
\subsection{A scalar field quintessence model}\label{scalar}
 One of the more interesting physical realizations of the dark energy is the so called quintessence, i.e. a self-interacting scalar field minimally coupled with gravity \citep[][]{Peebles88a, Peebles88b, Tsujikawa13}. Such field induces the repulsive gravitational force dynamically, driving the accelerated expansion of the Universe. Moreover, it also influences the growth of structures, arisen from gravitational instability. Quintessence could cluster gravitationally on very  large  scales  ($\simeq 100$  Mpc), and leaves an imprint on anisotropy of the  microwave  background  radiation  \citep[][]{cal+al98} , and, at  smaller scales, its fluctuations are damped and do not modify the evolution equation for the perturbations in the dark matter \citep[][]{ma+al99}.
Since redshift space distortions in the clustering of galaxies provide constrains on the growth rate of matter perturbations, $\delta_m$, which depend on the scalar field dynamics, and on the scalar field equation of state, it is possible to test quintessence models  from this kind of data \citep[see for instance][and references therein]{Alcaniz01,Copeland06, Bueno-Perivo}. Indeed in \citep[][]{MECC05} we showed that some scalar models with exponential potential, including the one used in the present analysis,  are  fully compatible with  the power spectrum of the CMBR anisotropy, and the parameters of large scale structure determined by the 2-degree Field Galaxy Redshift Survey (2dFGRS).  
It is worth noting that, in absence of the matter term, it is possible to deeply connect extended theories of gravity, as $f (R)$ theories and scalar tensor theories in the so called Jordan frame, to minimally coupled scalar field with appropriate self interaction potential. Actually, to any $f(R)$ theory in the so called Jordan frame, in the Einstein frame corresponds a minimally coupled scalar field, throughout the conformal transformation

\begin{displaymath}
g_{\mu \nu} \rightarrow \tilde{g}_{\mu \nu} = \Omega^2 g_{\mu \nu} \ ,
\end{displaymath}
where $\Omega \equiv \exp{(\varphi)} = \sqrt{f'(R)}$,  with a potential given by  \citep[][]{confpot}:

\begin{equation}
V(\tilde{\varphi}) = \frac{f(R) - R f'(R)}{2 \left [ f'(R) \right ]^2 } \,.
\label{eq: potconf}
\end{equation}
Here tilted quantities refer to the Einstein frame and  $\prime$ denotes partial derivative with respect to $R$. It is worth highlighting, however, that this connection only holds in the absence of a matter term. Indeed, even if it is possible to derive a conformal transformation in  presence of a matter term, the  self-interaction  potential  is not expressed  as  in Eq. (\ref{eq: potconf}), and matter becomes coupled to the scalar field. Therefore the physical equivalence of these two conformally {\it equivalent} theories remains an open question in theoretical cosmology \citep[see for instance][]{ma+al99,Cap06,MECP08}. Many quintessence
models have  been proposed, considering different kinds of
potentials driving the dynamics of the scalar field.  Here we
take into account the specific class of exponential--type
potential; in particular we consider an exponential potential
for which  general exact solutions of the Friedman equations are
known \citep[][]{PaoloClau02, MECC05, MEC11, EPGC}. Assuming that  $\varphi$ is
minimally coupled to gravity, the cosmological equations are written as
\begin{equation}
\left( \frac{\dot{a}}{a} \right)^{2} = \frac{8\pi G}{3}\left( Da^{-3} + \frac{1}{2}{\dot{\varphi}}^2 + V(\varphi)
\right)\,, \label{eq7}
\end{equation}
\begin{equation}
2\frac{\ddot{a}}{a} + \left( \frac{\dot{a}}{a} \right)^{2} = -\frac{8\pi G}{3}\left( \frac{1}{2}{\dot{\varphi}}^2 -
V(\varphi) \right)\,, \label{eq8}
\end{equation}
\begin{equation}
\ddot{\varphi} + 3\left( \frac{\dot{a}}{a} \right)\dot{\varphi} + V^{\prime}(\varphi) = 0\,, \label{eq9}
\end{equation}
and $\prime$ denotes partial derivative with respect to $\varphi$. Here we consider the potential analyzed in \citep[][]{MEC11} and \citep[][]{EPGC},
\begin{equation}\label{scal1}
V(\varphi) \propto \exp\left\{ -\sqrt{3\over 2}\varphi\right\}\,,
\end{equation}
for which the general exact solution exists: actually it turns out that
\begin{eqnarray}\label{eq:aHOMtime}
&&a^3(t)={t^2\over 2}[(3\mathcal{ H}_0-2)t^2+4-3\mathcal{ H}_0],\label{eq:aHOMtime1}\\
&&H(t)=
{{2\left(2(3\mathcal{ H}_0-2)t^2+4-3\mathcal{ H}_0\right)}\over{3t\left((3\mathcal{ H}_0-2)t^2+4-3\mathcal{ H}_0\right)}},\label{eq:aHOMtime2}\\
&&\Omega_M={{(4-3\mathcal{ H}_0)\left((3\mathcal{ H}_0-2)t^2+4-3\mathcal{ H}_0\right)}\over[2(3\mathcal{ H}_0-2)t^2+4-3\mathcal{ H}_0]^2},\label{eq:aHOMtime3}\\
&&\Omega_{\varphi}={{(3\mathcal{ H}_0-2)t^2\left(4(3\mathcal{ H}_0-2)t^2+3(4-3\mathcal{ H}_0)\right)}\over
[2(3\mathcal{ H}_0-2)t^2+4-3\mathcal{ H}_0]^2}\label{eq:aHOMtime4}\\
&&\varphi(t)=-\sqrt{\frac{2}{3}} \log \left(\frac{6.48}{\left(3 \mathcal{ H}_0-2\right) t^2-3 \mathcal{ H}_0+4}\right)\,,
\end{eqnarray}
where $\mathcal{H}_0$ is a constant.
In order to determine the integration
constants we set the present time $t_0 =1$, so we are using the
age of the universe as a unit of time, so $a_0 = a(1) = 1$, which
is a standard choice, and finally $ \mathcal{ H}_0= H(1)$.
Because of our choice of time unit $\mathcal{H}_0$ does not
have the same value as  the standard Hubble constant $H_0$.
In this model all the basic cosmological parameters can be written in terms of $\mathcal{H}_0$ only,
so we find that:
\begin{equation}\label{eq:om_matternow}
\Omega_{M_0}\equiv\Omega_{M}(t=1)=\frac{2(4 -
3\,\mathcal{ H}_0)}{9\,\mathcal{ H}_0^2}\,,
\end{equation}
\begin{equation}\label{eq:omphinow}
\Omega_{\varphi_0}\equiv\Omega_{\varphi}(t=1)= \frac{\left(3\,\mathcal{ H}_0
-
  2\right)
\,\left(3 \,\mathcal{ H}_0 + 4 \right) }{9\,\mathcal{ H}_0^2}\,.
\end{equation}

\begin{figure}
 \begin{minipage}[b]{0.9\linewidth}
\centerline{ \includegraphics[width=.7\linewidth,height=.7\linewidth]{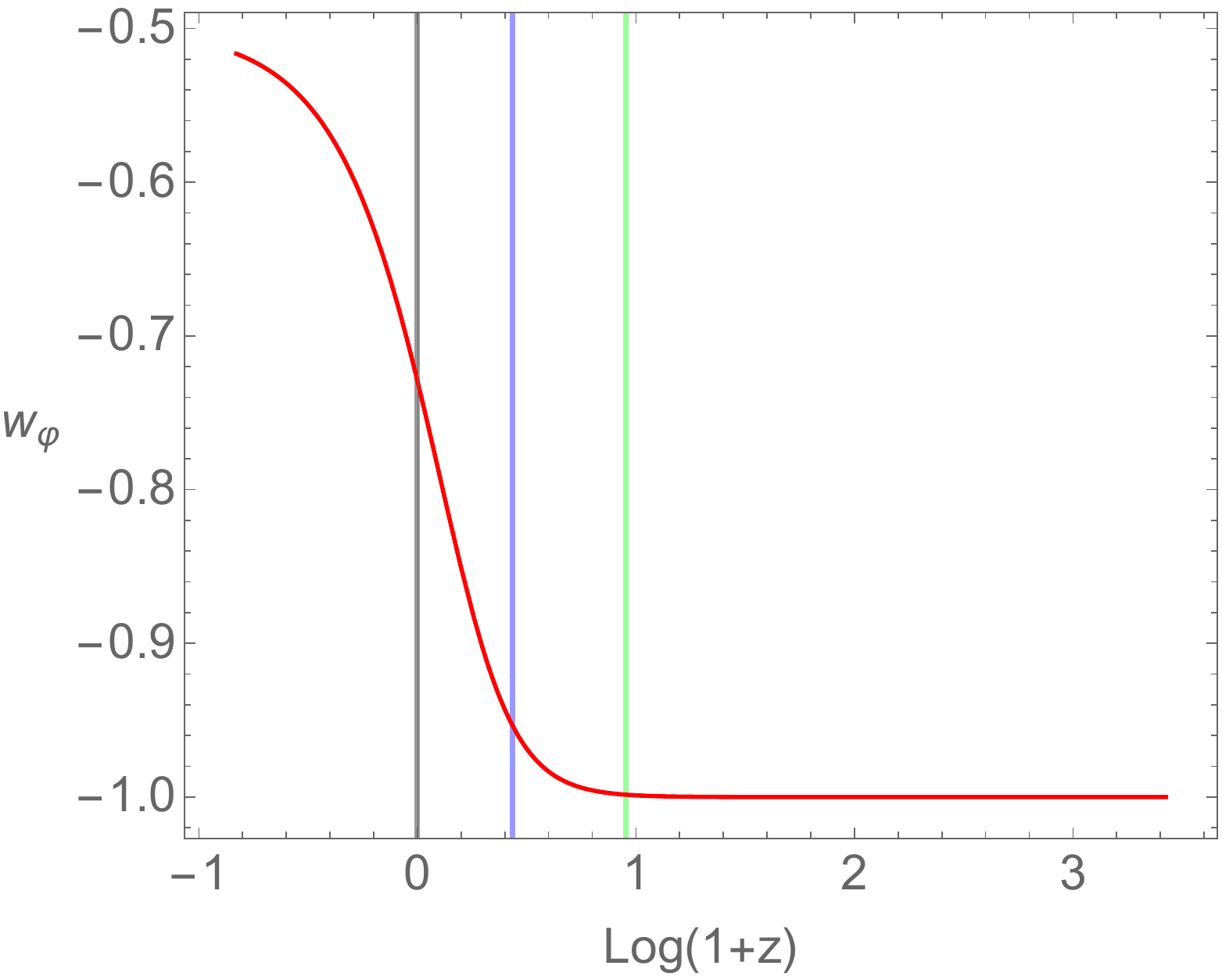}}
  \end{minipage}
\caption{
Redshift dependence of the equation of state parameter $w_{\varphi}$ corresponding to the best fit value of $H_0$, as illustrated in the statistical analysis section: we see that $w_{\varphi}$ smoothly transits between two asymptotic constant values. The vertical lines indicate respectively the SNIa (blue line) and the GRB range of redshift (green line).}
\label{wphi}
\end{figure}

\begin{figure}
 \begin{minipage}[b]{0.9\linewidth}
\centerline{ \includegraphics[width=.7\linewidth,height=.7\linewidth]{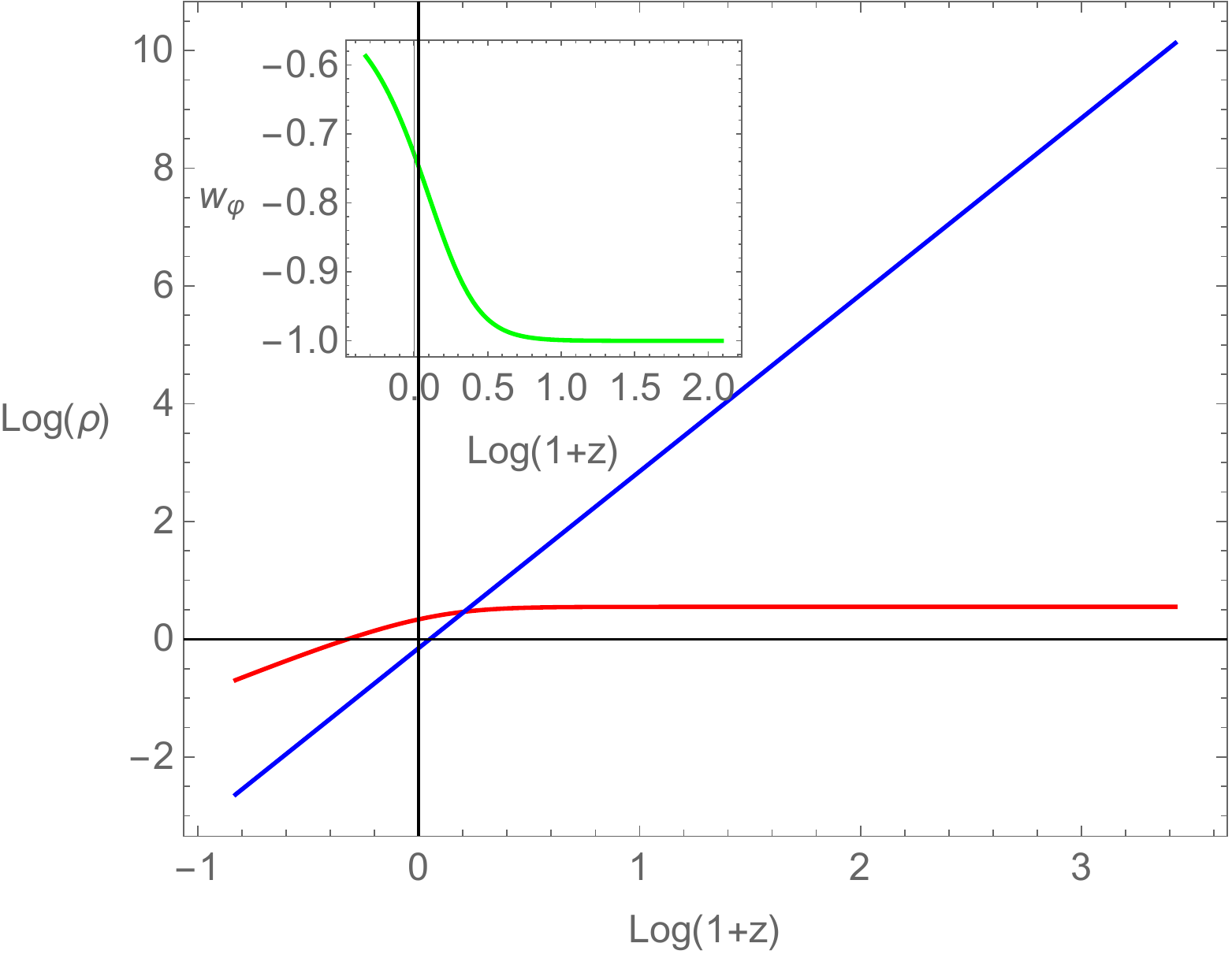}}
  \end{minipage}
\caption{
The scalar field density energy $\rho_{\varphi}$ (red line) is compared with the matter density (blue line): we see the transition from a  subdominant phase, during the matter-dominated era, to a dominant phase, at the present accelerated expansion. It turns out that the range of redshifts of this transition corresponds to the passage of the equation of state from $w_{\varphi}= -1$ to its present value.   }

\label{logrho}
\end{figure}

The  scalar field EOS evolves with time and the
parameter $w$ is given by
\begin{equation}\label{w}
w_{\varphi}= -\frac{1}{2}  +
  \frac{ 3(3\,\mathcal{ H}_0 - 4) }{6(4 -
     3\,\mathcal{ H}_0) +
     8\left(3\,\mathcal{ H}_0 - 2
                \right)t^2}\,,
        \end{equation}
so that today we have
\begin{equation}\label{w0}
 w_{\varphi_0}= -{8-3 \mathcal{ H}_0\over 4+3 \mathcal{ H}_0}\,.
\end{equation}
As we see in Fig. (\ref{wphi} ), the scalar field EOS in the past is equal to $w_{\varphi}=-1$ so that the dark energy behaves as a subdominant cosmological constant, but only recently has started dominating the expansion of the universe, as illustrated in Fig. (\ref{logrho}). It turns out that it undergoes a transition from a subdominant phase, during the matter-dominated era, to a dominant phase, associated to the present accelerated expansion. It is clear that  extending the Hubble diagram beyond the SNIa range of redshift allows us to fully investigate the transition of the EOS from $w_{\phi}=-1$ to its present value. 

\subsection{Early dark energy}\label{ede}
\begin{figure}
  \begin{minipage}[b]{0.95\linewidth}
\includegraphics[width=.95\linewidth,height=.8\linewidth]{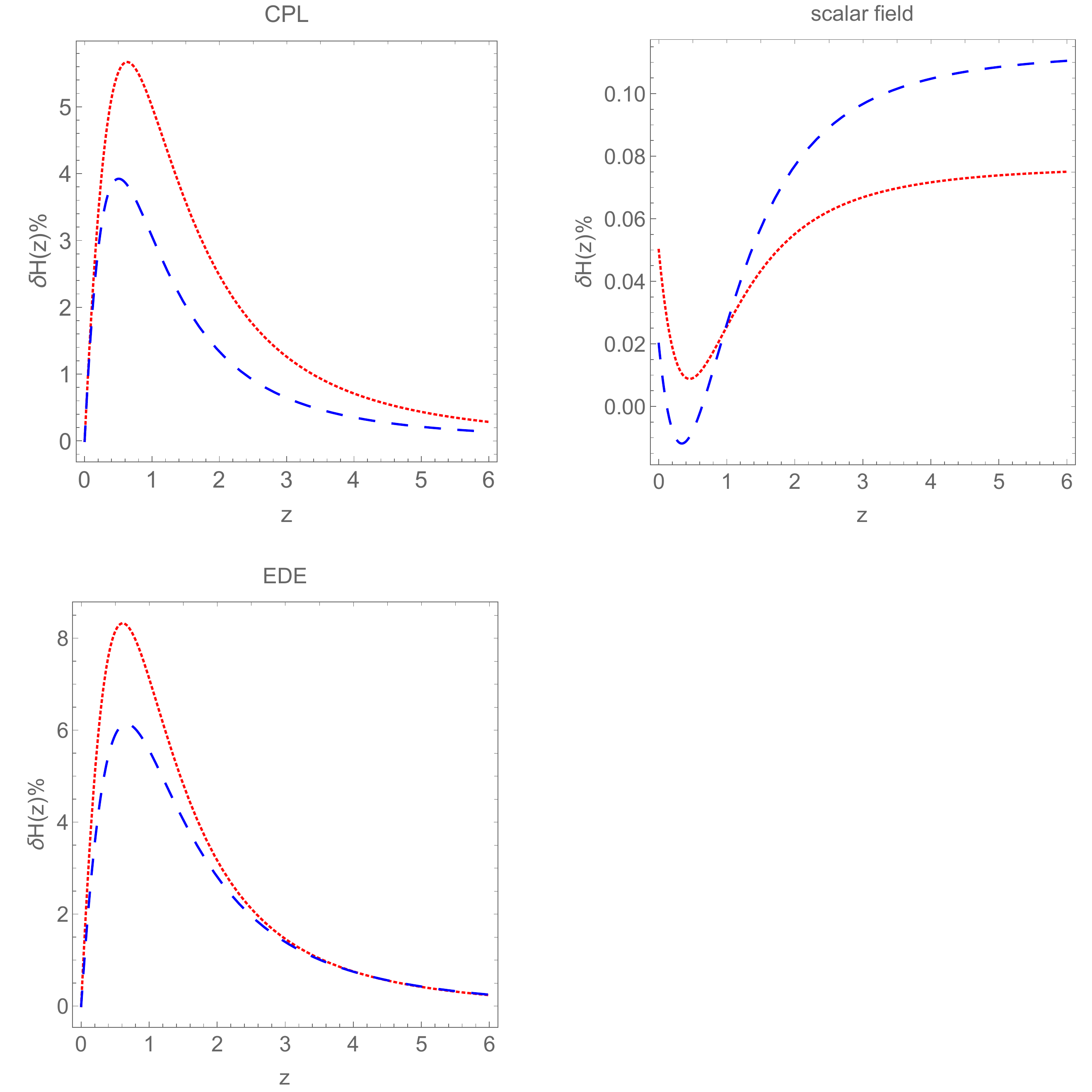}
\end{minipage}
\caption{We show relative residual curves of $\displaystyle\frac{H(z,\Lambda)-H(z,\mathbf{\theta})}{ H(z,\Lambda)}$ for all considered competing models described above. For the flat $\Lambda$CDM model we take $\Omega_m=0.3$, and $H_0=70$. For other models the values of the parameters have been chosen in order to highlight the differences between models. In particular, for the CPL  model $\Omega_m=0.3$, $h=0.7$, $w_0=-1.2$, $w_1=0.2$ (red line) and $\Omega_m=0.3$\,,$h=0.7$, $w_0=-1.2$, $w_1=-0.2$ (blue line); for the EDE model $\Omega_m=0.3$, $h=0.7$, $w_0=-1$, $\Omega_e=0.08$, $\omega_{\gamma}= 2.47 \,\, 10^{-5}$ (red line), $\Omega_m=0.27$, $h=0.7$, $w_0=-1$, $\Omega_e=0.03$, $\omega_{\gamma}= 2.47 \,\, 10^{-5}$(blue line). For the scalar field model $\mathcal{ H}_0 =0.94$ (red line) and $\mathcal{ H}_0 =0.98$. } \label{pH}
\end{figure}

\begin{figure}
 \begin{minipage}[b]{0.95\linewidth}
\centerline{ \includegraphics[width=.7\linewidth,height=.7\linewidth]{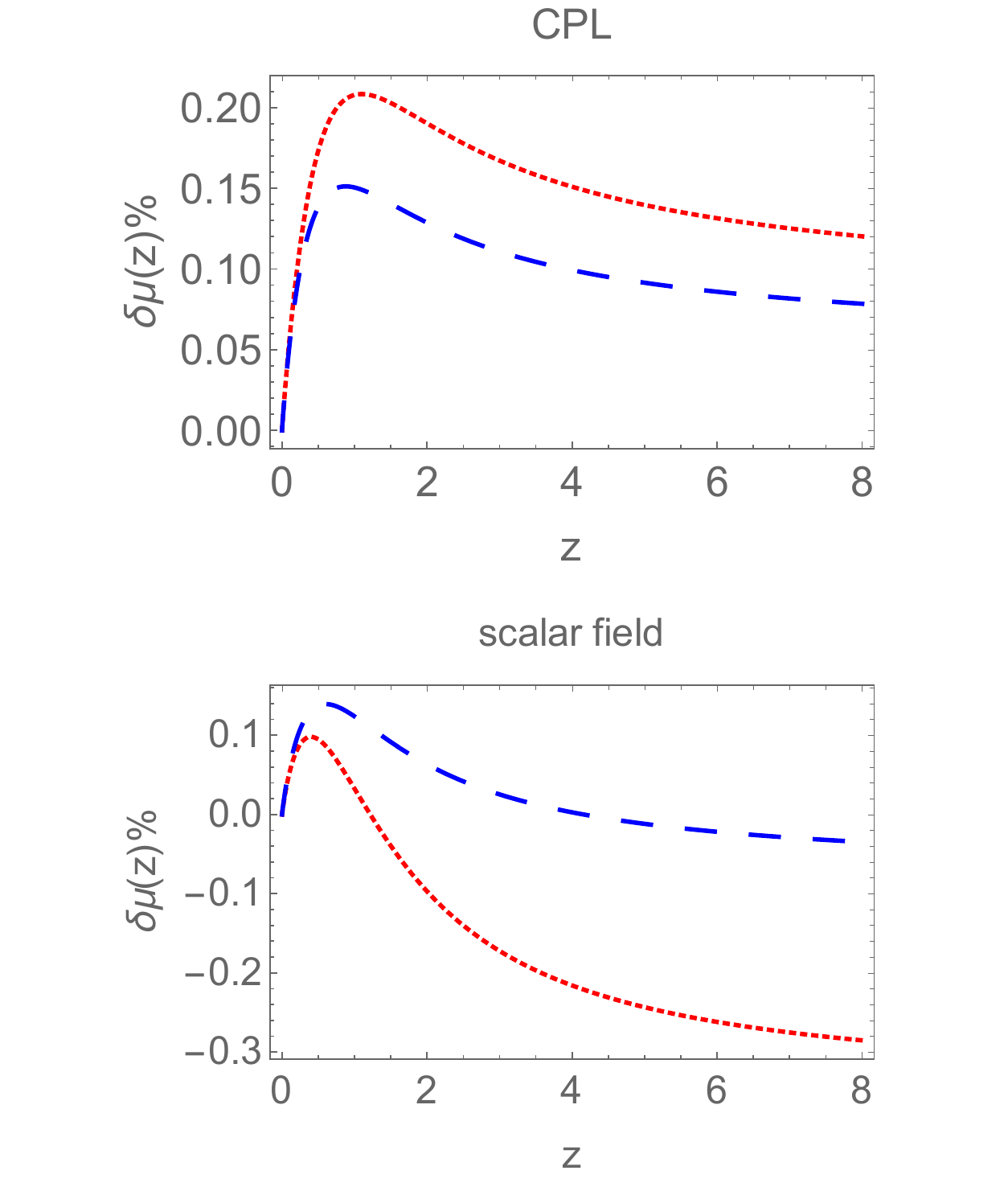}}
  \end{minipage}
 \begin{minipage}[b]{0.95\linewidth}
\centerline{\includegraphics[width=.6\linewidth, height=.5\linewidth]{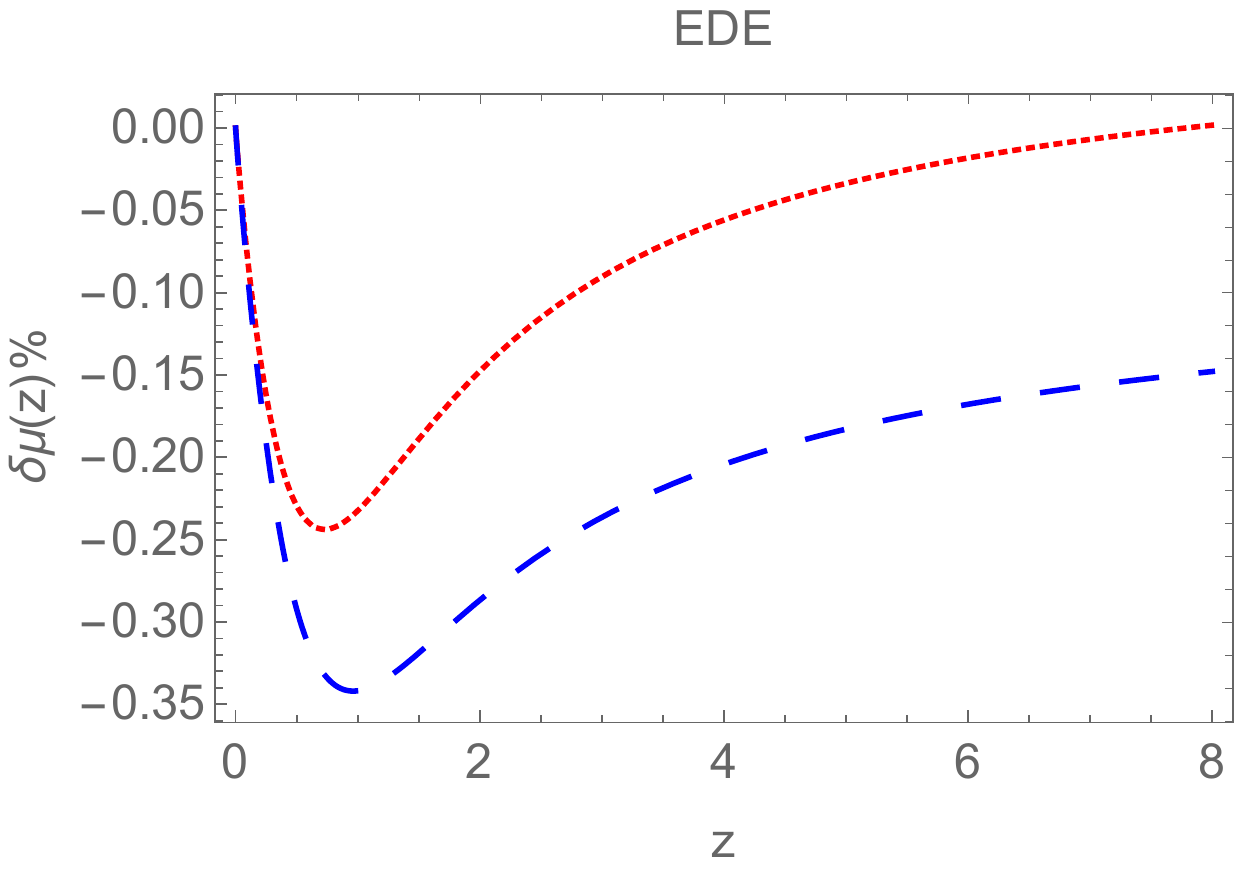}}
  \end{minipage}
\caption{We show relative residual curves of $\displaystyle\frac{\mu(z,\Lambda)-\mu(z,\mathbf{\theta})}{ \mu(z,\Lambda)}$ for all our considered models described above. For the flat $\Lambda$CDM model we take $\Omega_m=0.3$, and $H_0=70$.  For other models the values of parameters have been chosen as in Fig.(\ref{pH}).
 For the EDE model we take $\Omega_m=0.3$, $h=0.7$, $w_0=-1$, $\Omega_e=0.08$, $\omega_{\gamma}= 2.47 \,\, 10^{-5}$ (red line), and $\Omega_m=0.27$, $h=0.7$, $w_0=-1$, $\Omega_e=0.03$, $\omega_{\gamma}= 2.47 \,\, 10^{-5}$(blue line). } \label{DeltaMu-Muede}
\end{figure}

In this section we consider some proposed cosmological models
that allow a non negligible amount of dark energy at early
times \citep[][]{doran06}:  they are often connected with the existence
of scaling  or attractor-like solutions, in which the dark energy
density follows the density of the dominant component of matter-energy
in the Universe. These
models naturally predict a non-vanishing dark energy fraction
of the total energy at early times, $\Omega_{e}$, which should
be substantially smaller than its present value. Therefore these models need an {\it exit mechanism}, allowing the scaling solutions to end in the recent cosmological past, in order to trigger a dark energy dominant era.  A large class
of models of this type has been proposed \cite[][]{Karwal16, Niedermann}. Since the main parameter of an early dark energy model is  $\Omega_{e}$, it parametrizes  the evolution of dark energy. In different parameterizations $\Omega_{e}$ have been estimated from several observations, as nucleosynthesis, structure formation, or the peak separation  in the power spectrum of the cosmic microwave radiation anisotropy\citep[][]{das,doran02,doran06,divalentino21}.  Following
\citep[][]{doran06, pettorino13} we use  parametrized representation
of the  dark energy density fraction, $\Omega_{de}$, which
depends on the present matter fraction, $\Omega_m$, the early
dark energy density fraction, $\Omega_e$ , and the present dark
energy equation of state $w_0$:
\begin{eqnarray}\label{edep}
&&\Omega_{de}(z,\Omega_m, \Omega_e, w_0)=\frac{\Omega _e \left((z+1)^{3 w_0}-1\right)-\Omega _m+1}{\Omega _m(z+1)^{-3 w_0}-\Omega _m+1}\, +\nonumber\\
   && + \Omega _e \left(1-(z+1)^{3 w_0}\right)\,.
\end{eqnarray}
It turns out that in these models the Hubble function takes the form:
 \begin{eqnarray}\label{Hede}
&&H^{2}(z,\Omega_m, \Omega_e, w_0,\Omega _{\gamma }, N_{eff })=\Omega_{de}(z,\Omega_m, \Omega_e, w_0)+\nonumber \\
&&+ \Omega _m (z+1)^3+ \Omega _{\gamma } (z+1)^4 \left( \frac{7}{8}\left(\frac{4}{11}\right)^{\frac{4}{3}} N_{eff }+1\right).
 \end{eqnarray}
Here $N_{eff }$ is defined so that the total relativistic energy density (including neutrinos and any other dark radiation)
is given in terms of the photon density $\rho_{\gamma}$ at $T \ll1$ MeV by the relation:
\begin{equation}
\rho=N_{eff } \frac{7}{8}\left(\frac{4}{11}\right)^{\frac{4}{3}}\rho_{\gamma}\,.
 \end{equation}
In this equation $N_{eff}=3$ for three standard neutrinos that
were thermalized in the early Universe and decoupled well before
electron-positron annihilation. Moreover $\Omega_\gamma=\omega_\gamma h^{-2}$, and we set $\omega_\gamma= 2.47 \,\, 10^{-5}$.
In Figs. (\ref{pH}) and (\ref{DeltaMu-Muede}) we plot relative residual curves of $\displaystyle\frac{H(z,\Lambda)-H(z,\mathbf{\theta})}{ H(z,\Lambda)}$, and $\displaystyle\frac{\mu(z,\Lambda)-\mu(z,\mathbf{\theta})}{ \mu(z,\Lambda)}$, for all the competing models described above, where $H(z,\Lambda)$ is the Hubble function in the standard flat $\Lambda$CDM model, and $H(z,\mathbf{\theta})$ is the Hubble function in each of the other models parametrized by $\mathbf{\theta}$. We plot also curves of $\frac{\mu(z,\Lambda)-\mu(z, {\mathbf{\theta}}) }{ \mu(z,\Lambda)}$ for the CPL model and scalar field models: it turns out that the range of redshifts larger than $z=1$ is very important to break degeneracies among the flat standard $\Lambda$CDM and other models considered in this paper.  The values of the parameters have been chosen in order to highlight the differences among models.

\section{Observational data}
In our analysis we use the  SNIa and
GRB Hubble diagrams, and a list of
$28$ direct $H(z)$ measurements, compiled by \citep[][]{farooqb}.

\subsection{Supernovae}
SNIa observations gave the first strong indication that now the expansion rate of the Universe is accelerating.
First results of the SNIa teams were
published in  \citep[][]{Riess} and \citep[][]{per+al99}. Here we consider
the recently updated Supernovae Cosmology Project Union 2.1
compilation \citep{Union2.1}, which is an update of the original
Union compilation. The Union2.1 dataset is one of the largest compilations of supernovae of type Ia,  originally consisting of 833 SNIe drawn from different samples, and reduced to 580  after several selection criteria. It spans the redshift
range $0.015 \le z \le 1.4$. To compute the $\chi^2$ function related to the distance modulus we can easily relate the
apparent magnitude $m(z)$ to
the so called Hubble free luminosity distance $D_L(z)= H_0 d_L(z)$ through  the relation:
\begin{equation}
m_{th}(z)={\bar M}+ 5 \log_{10}D_L (z)\,. \label{mdl}
\end{equation}
Here ${\bar M}$ is the zero point offset, it depends on the absolute magnitude
$M$ and on the present value of the Hubble parameter $H_0$:
 \begin{eqnarray}
{\bar M} &=& M-5\log_{10}h+42.38\,,\label{barm}
\end{eqnarray}
where $M$  is the absolute magnitude. The
cosmological model parameters can be determined by minimizing
the quantity
\begin{equation}
\chi^2_{SNIa} ( \{\theta_{p}\})= \sum_{i=1}^N \frac{(\mu_{obs}(z_i) -
\mu_{th}(z_i,\{\theta_{p}\}))^2}{\sigma_{\mu \; i}^2 }\,.\label{chi2}
\end{equation}
Here $\sigma_{\mu \; i}^2 = \bar{\sigma}_{\mu \; i}^2+\sigma_{int}^2$, where $\sigma_{int}$ is a fit parameter; in our statistical analysis we will marginalize over $\sigma_{int}$.
The theoretical distance modulus is defined as
\begin{equation}
\mu_{th}(z_i,\{\theta_{p}\}) =5 \log_{10} D_L
(z_i,\{\theta_{p}\}) +\nu_0\,, \label{mth}
\end{equation}
where  $\nu_0= 42.38 - 5 \log_{10}h$, and  $\{\theta_{p}\}$
denotes the set of parameters that appear in different dark
energy models \citep[][]{Nesseris05}. For example, in the case of a
flat CPL model $\{\theta_{p}\}=\{\Omega_m, w_0, w_1$\}.
\subsection{Gamma-ray bursts}
Gamma-ray bursts are visible up to high redshifts thanks to the
enormous energy that they release, and thus may be good
candidates  for our high-redshift cosmological investigation \citep[see, for instance,][]{Lin15, Lin16, Lin16b,Dainotti,AmatiDisha,Izzo,Wei17,Si18,Fana19,Khadka20,Zhao20}, or \citep[][]{ZhaoApJ, Cao,muccino21b,Khadka20}. However, GRBs may be everything but
standard candles since their peak luminosity spans a wide range,
even if  there have been many efforts to make them distance
indicators using some empirical correlations of
distance-dependent quantities and rest-frame observables
\citep[][]{Amati08}.

\begin{table}
\begin{center}
\resizebox{12 cm}{!}{\begin{minipage}{\textwidth}
\begin{tabular}{ cccccc }
\,& \,& \multicolumn{1}{c}{\bf Dependence on redshift bins}   \\
\, & \, & \, & \, & \, &  \\
\hline
\multicolumn{1}{c}{\textbf{E$ _{p,i}$$-$E$ _{iso}$ Correlation}} & {\textbf{total GRBs}} & {\textbf{normalization}} & {\textbf{slope}} & {\textbf{scatter}} \\
 \hline
\hline\\

z $<$ 0.5 & 13 & 1.97$\pm$0.07 & 0.60$\pm$0.06 & 0.201$\pm$0.052\\
0.5 $<$ z $<$ 1 & 38 & 2.05$\pm$0.05 & 0.51$\pm$0.07 & 0.238$\pm$0.032\\
1 $<$ z $<$ 1.5 & 38 & 2.02$\pm$0.06 & 0.50$\pm$0.05 & 0.172$\pm$0.025\\
1.5 $<$ z $<$ 2 & 28 & 1.95$\pm$0.16 & 0.56$\pm$0.11 & 0.230$\pm$0.041\\
2 $<$ z $<$ 2.5 & 28 & 2.13$\pm$0.11 & 0.44$\pm$0.07& 0.174$\pm$0.031\\
2.5 $<$ z $<$ 3 & 16 & 1.78$\pm$0.11 & 0.68$\pm$0.08 & 0.101$\pm$0.031\\
3 $<$ z $<$ 3.5 & 13 & 1.99$\pm$0.13 & 0.55$\pm$0.10 & 0.135$\pm$0.048\\
3.5 $<$ z $<$ 4 & 7 & 2.16$\pm$0.15 & 0.44$\pm$0.10 & 0.069$\pm$0.008\\
4 $<$ z  & 12 &  2.06$\pm$0.13 & 0.51$\pm$0.11 & 0.116$\pm$0.058\\
\hline
\end{tabular}
\end{minipage}}
\caption{Dependence of the E$ _{p,i}$$-$E$ _{iso}$
correlation on different redshift bins.}
\label{tabep-eisoz}
\end{center}
\end{table}

 Actually GRBs show non thermal spectra which can be empirically modeled with the  Band function \citep[][]{Band}, which is a smoothly broken power law with parameters  $\alpha$, the low-energy spectral index, $\gamma$ , the high energy spectral index and the {\it roll-over} energy $E_0$. Their spectra show a peak corresponding to a value of the photon energy $E_{\rm p} = E_0 (2 + \alpha)$; indeed it turns out that for GRBs with measured spectrum and redshift it is possible to evaluate the intrinsic peak energy, $E_{\rm p,i} = E_{\rm p} (1 + z)$ and the isotropic equivalent radiated energy, defined as:
\begin{equation}
E_{\rm iso}= 4 \pi d_L^2(z,{\mathrm \theta}) \left(1+z\right)^{-1}\int^{10^4/(1+z)}_{1/(1+z)} E N(E)
dE\,.
\label{eqEiso}
\end{equation}

Here $N(E)$ is the Band function:
\[N(E)=\left\{
\begin{array}{ll}
 A \left(\frac{E}{100keV}\right)^{\alpha_B} \exp{\left(-{\frac{E}{E_0}}\right)}\,,\left(\alpha_B\right)E_0\geq E\,,\\
 A \left(\frac{\left(\alpha_B-\beta_B\right)E}{100keV}\right)^{\alpha_B-\beta_B} \exp{\left(\alpha_B-\beta_B\right)\left(\frac{E}{100keV}\right)^{\beta_B}}\,,\\ \left(\alpha_B-\beta_B\right)E_0\leq E\,.\\
\end{array}
\right. \]

Even if  $E_{\rm p,i} $ and $E_{\rm iso} $ span several orders of magnitude, it turned out that they are strongly correlated, according to the relation \citep[][]{Amati02}.

 \begin{equation}
 \log \left(\frac{E_{\rm iso}}{1\;\mathrm{erg}}\right)=b+a \log \left[\frac{E_{\mathrm{p, i}} (1+z)}{300\;\mathrm{keV}}\right]\,,
 \label{correlation}
\end{equation}
where $a$ and $b$ are constants. This correlation, as other correlations, is characterized by an extra-Poissonian scatter, $\sigma_{int}$,
distributed around the best fit law. It is clear from Eqs. (\ref{eqEiso} and \ref{correlation}) that it is possible to use GRBs to investigate cosmological models, if we can calibrate the $E_{\rm p,i} $ - $E_{\rm iso} $ correlation in a model independent way, overcoming the so called circularity-problem, which affects the estimation of the luminosity distance from all the GBR correlations. Actually it turns out that GRBs can be used as cosmological tools through the $E_{\rm p ,i}$ -- $E_{\rm iso}$ correlation; however the computation of $E_{\rm iso}$ is based on a fiducial cosmological model. Different and alternative techniques have been recently developed in literature  \citep[see for instance][]{montiel,muccino19,muccino21,Izzo,Wang15, Liang08,Kodama08,Wei10,Lin15}): here we standardize our GRB dataset updating a method previously adopted, as discussed in a later section.
The $E_{\rm p,i}$ -- $E_{\rm iso}$ data sample used in this analysis was build up  by  Amati and Sawant, they  collected the spectral information of GRBs with measured redshift from February 1997 to October 2015 \citep[][]{AmatiDisha}. This database includes redshift z, both energy indices $\alpha$ and $\gamma$, the peak energy $E_{\rm p,i}$ computed from the break energy $E_0$, $t_{90}$, exposure time, the fluence and the value of peak flux. The redshift distribution covers a broad range $0.033 \leq z < 9.0$, thus extending far beyond that of type Ia SN $z \leq 1.7$. For the oldest GRBs (BeppoSAX, BATSE, HETE-2) and other GRBs up to mid 2008, the data was adapted from \citep[][]{Amati08}. As already discussed in \citep[][]{MGRB1}, the criteria behind selecting the measurements from a particular mission are based on
the following conditions:
\begin{itemize}
\item[1.] We concentrated on observations for which the exposure time was at least 2/3rd of the whole event duration.
\item[2.] Given the broad energy band and good calibration, Konus-WIND and Fermi/GBM were chosen whenever available. For Konus-WIND, the measurements were taken from the official catalog \citep[][]{Ulanov05} and from GCN archives ($http://gcn.gsfc.nasa.gov/gcn3-circulars.html$). In the case of Fermi/GBM, the observations were derived from
\citep[][]{Gruber 12} and from several other papers, as, for instance, \citep[][]{Ghirlanda04}. The observations from SUZAKU were not considered as the uncertainties in the calibration are higher and also because it works in a narrow energy band.
\item[3.] The SWIFT BAT observations were chosen when no other preferred missions (Konus-WIND, Fermi/GBM) were able to provide information. They were considered only for  GRBs with the value of  $E_{\rm p,obs}$ that  was within the energy band of the instrument. For Swift GRBs, the  $E_{\rm p,i}$  value derived from BAT spectral analysis alone were conservatively taken from the results reported by the BAT team \citep[][]{Sakamoto08}. The GCN circulars were also used when needed.
\end{itemize}
When  more than one mission provides good observations based on these criteria,  the values and uncertainties of all those observations (hence more than one set for some finely observed GRBs) are taken into account. When the observations were to be included in the data sample, it has been checked  that the uncertainty on any value is not below $10\%$ in order to account for the instrumental capabilities, etc. So, when the error was lower, it has been assumed to be $10\%$.
When available, the Band model \citep{Band} was considered since the cut-off power law tends to overestimate the value of $E_{\rm p,i}$.
GRBs have been observed by different detectors, that are characterized by different thresholds
 and spectroscopic sensitivity, therefore they can spread relevant selection biases in the observed
 correlation. This is ongoing debated topic: in the past, there were claims that a large fraction
 $(70- 90\%)$ of BATSE GRBs without redshift is inconsistent with the correlation for any
 redshift \citep{Band05,Piran05}.  However other authors (\citep{Ghirlanda08,Nava11}) showed that, in fact,
  most BATSE GRBs with unknown redshift were consistent with the $E_{\rm p,i}$ -- $E_{\rm iso}$ correlation. We
  also note that inconsistency of a high percentage of GRBs of unknown redshift would  imply
  that most GRBs with known redshift should also be inconsistent with the $E_{\rm p,i}$ -- $E_{\rm iso}$  relation,
  and this fact was never observed. Moreover, \citep{Amati09} showed that the normalization of the correlation varies
  only marginally for GRBs observed by different instruments with different sensitivities and energy bands,
  while in other papers as, for instance, \citep{Ghirlanda10} and \citep[][]{AmatiDisha} it is shown that the parameters of the correlations  are independent of redshift.
 If the whole GRB sample is divided  into
redshift bins, as shown in Table \ref{tabep-eisoz} and Fig. (\ref{tabep-eisoz}), it turned out that the possible evolutionary effects are within the intrinsic scatter and, therefore, do not affect the correlation. Similar results were obtained in \citep[][]{A-D, MGRB1}, and \citep[][]{A-DV, Megrb11}

\begin{figure}
  \begin{minipage}[b]{0.5\linewidth}
   \includegraphics[width=.85\linewidth]{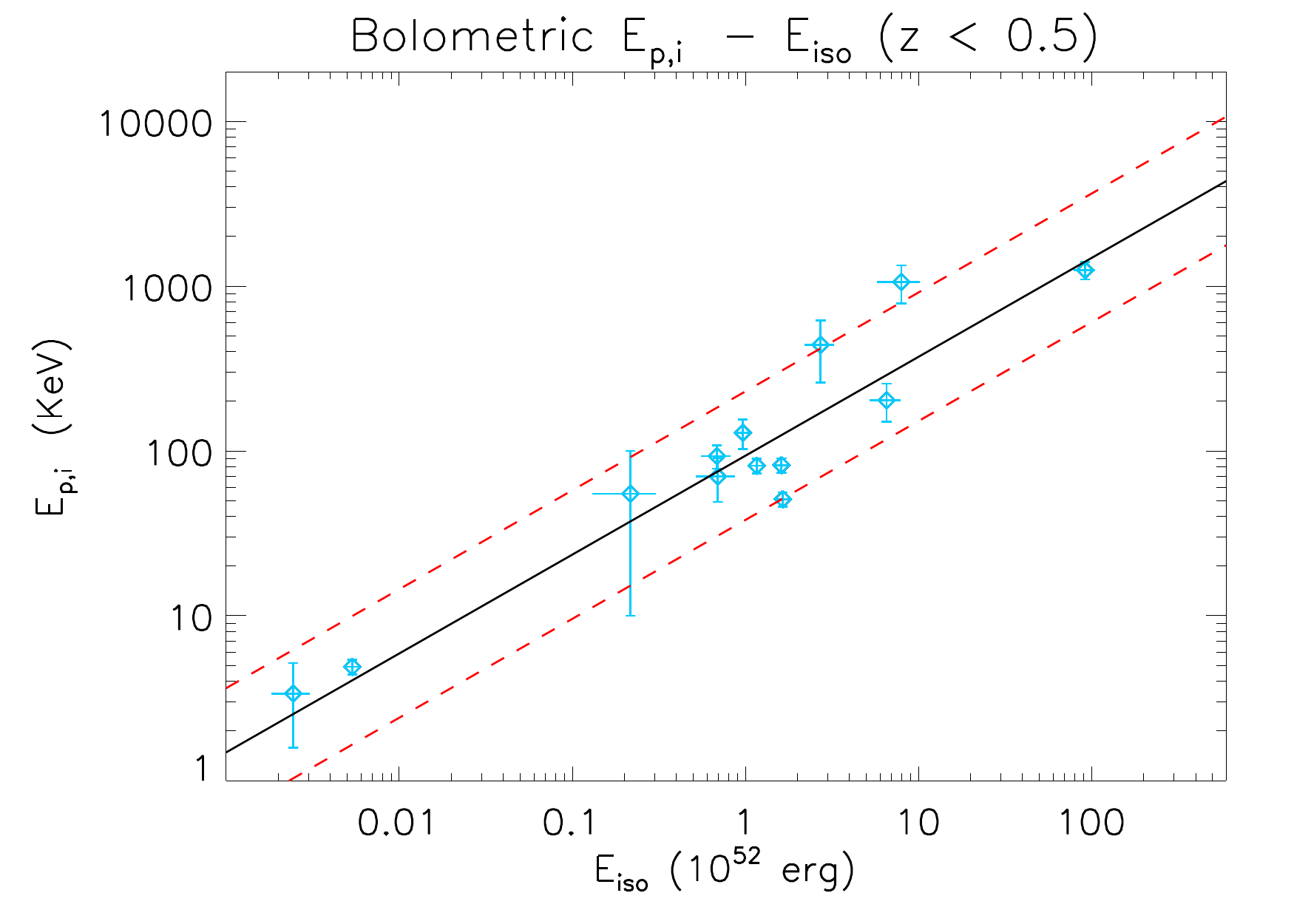}
  \end{minipage}
    \begin{minipage}[b]{0.5\linewidth}
    \includegraphics[width=.85\linewidth]{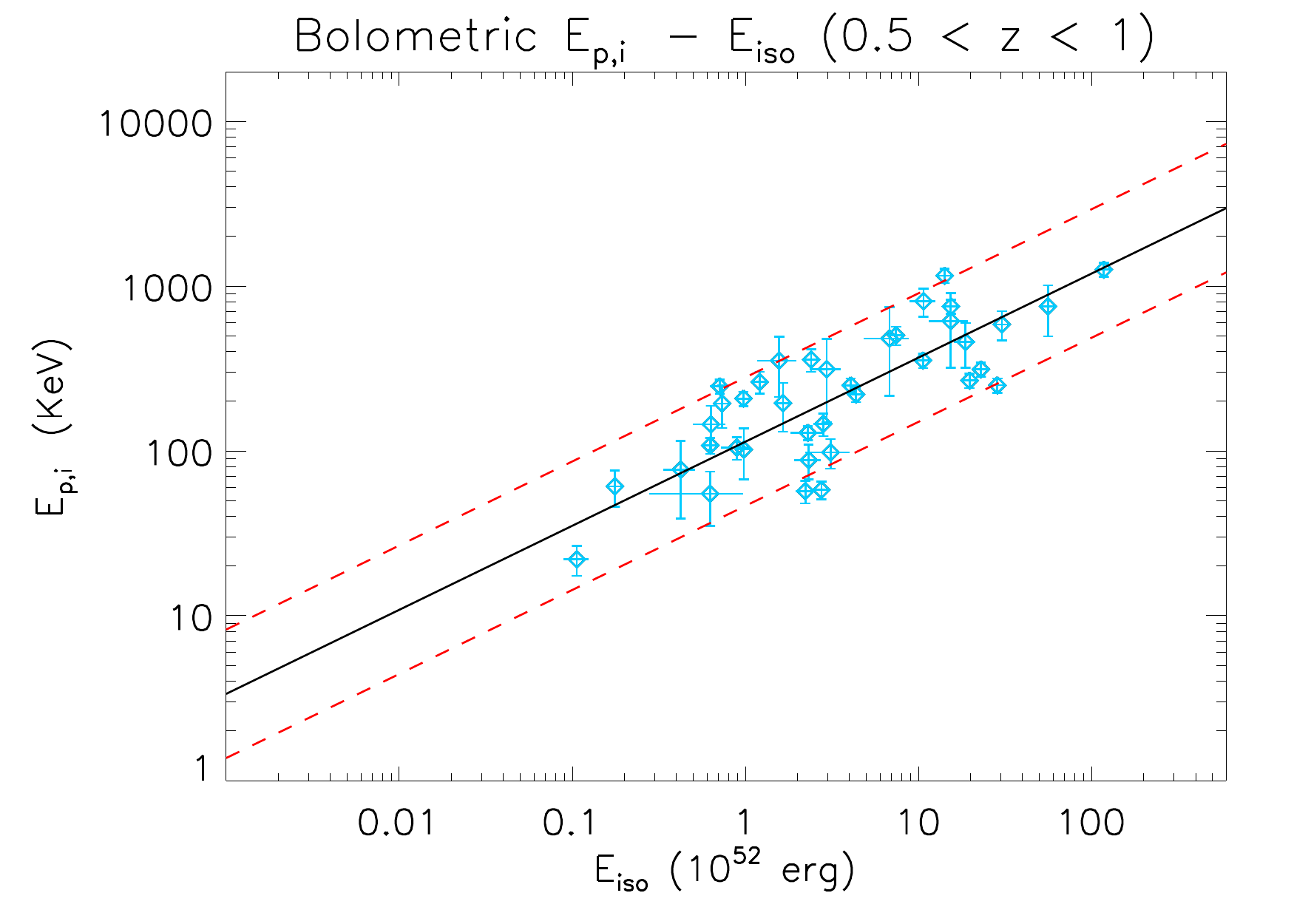}
  \end{minipage}
  \begin{minipage}[b]{0.5\linewidth}
   \includegraphics[width=.85\linewidth]{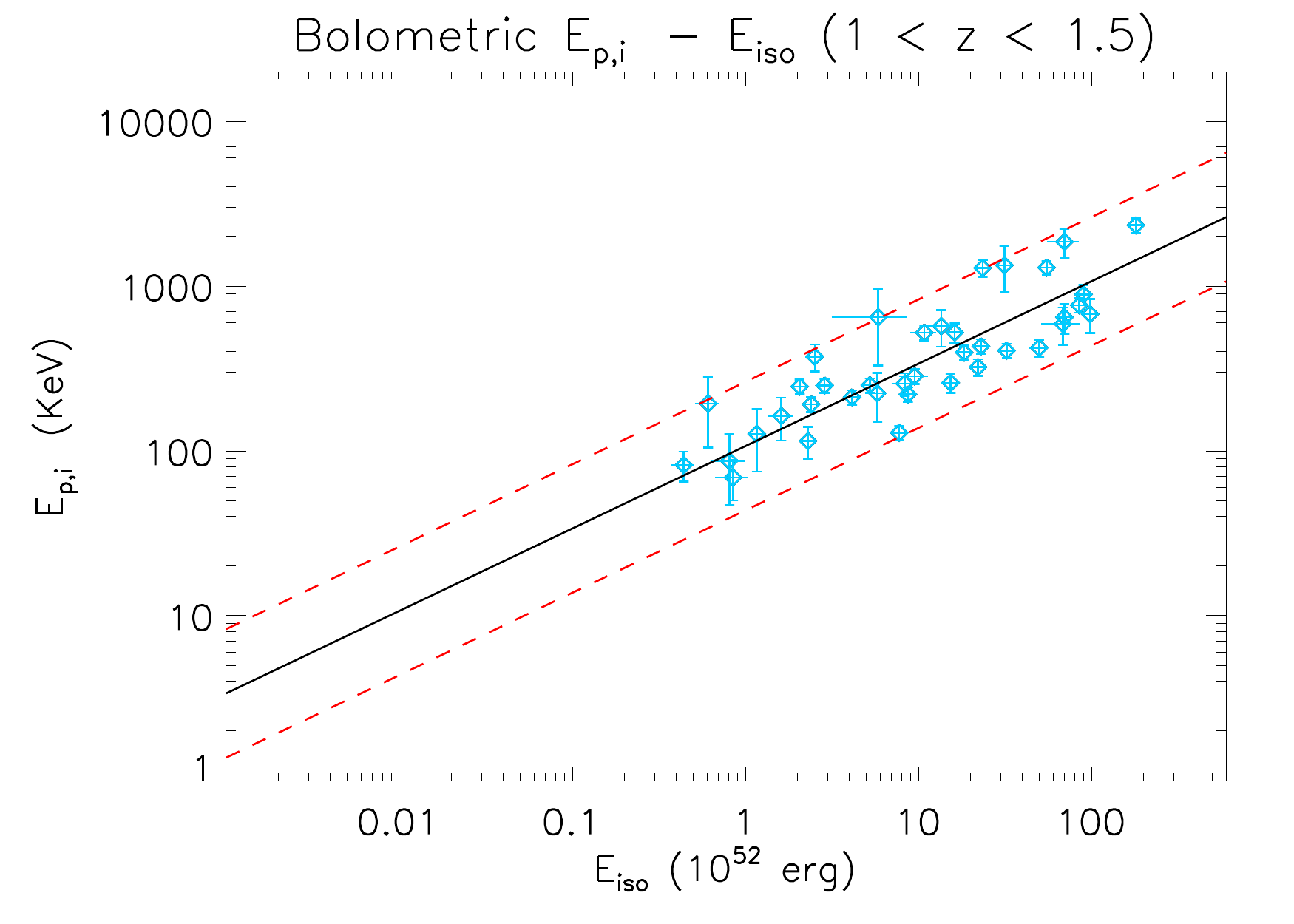}
  \end{minipage}
  \begin{minipage}[b]{0.5\linewidth}
   \includegraphics[width=.85\linewidth]{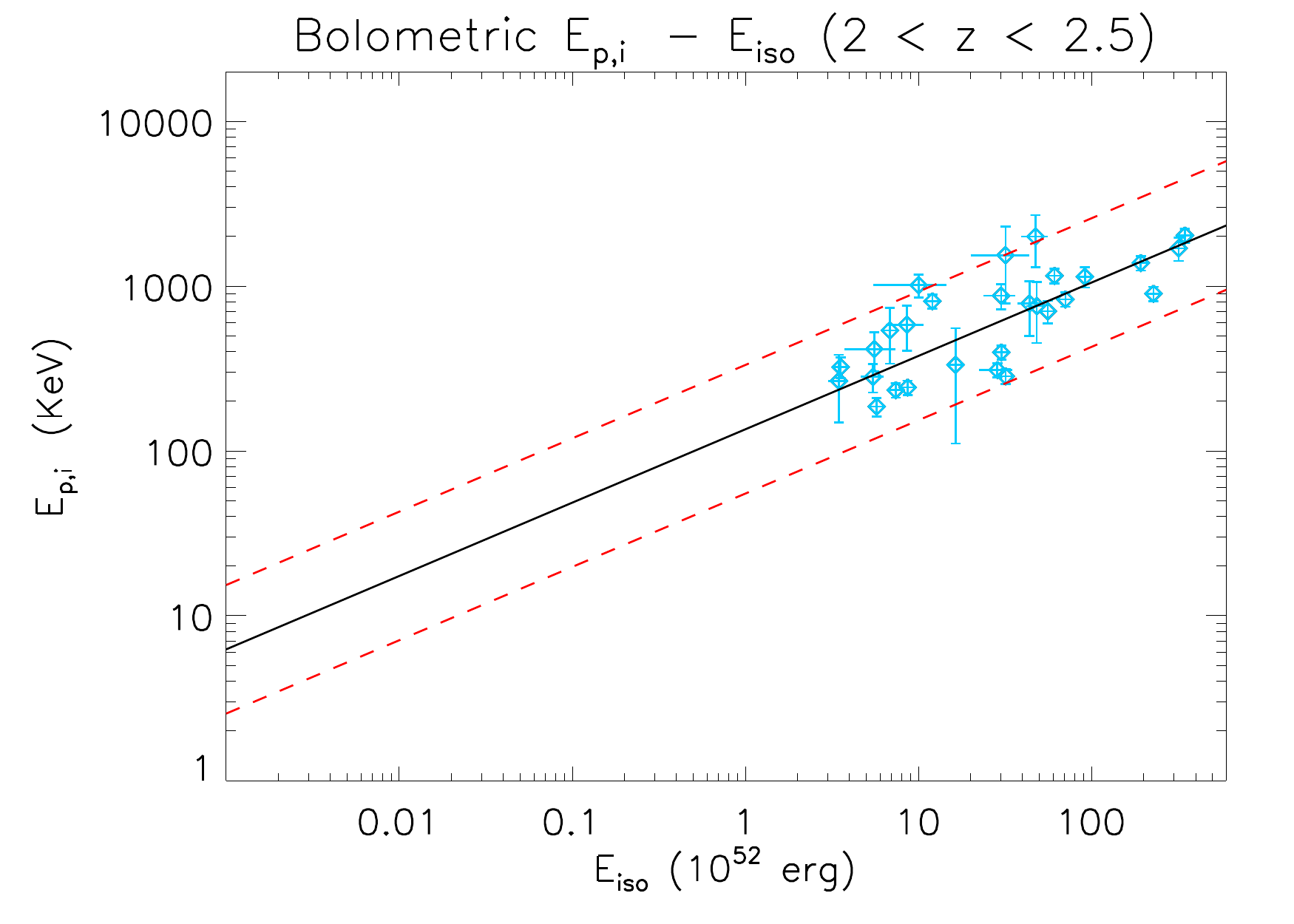}
  \end{minipage}
  \begin{minipage}[b]{0.5\linewidth}
   \includegraphics[width=.85\linewidth]{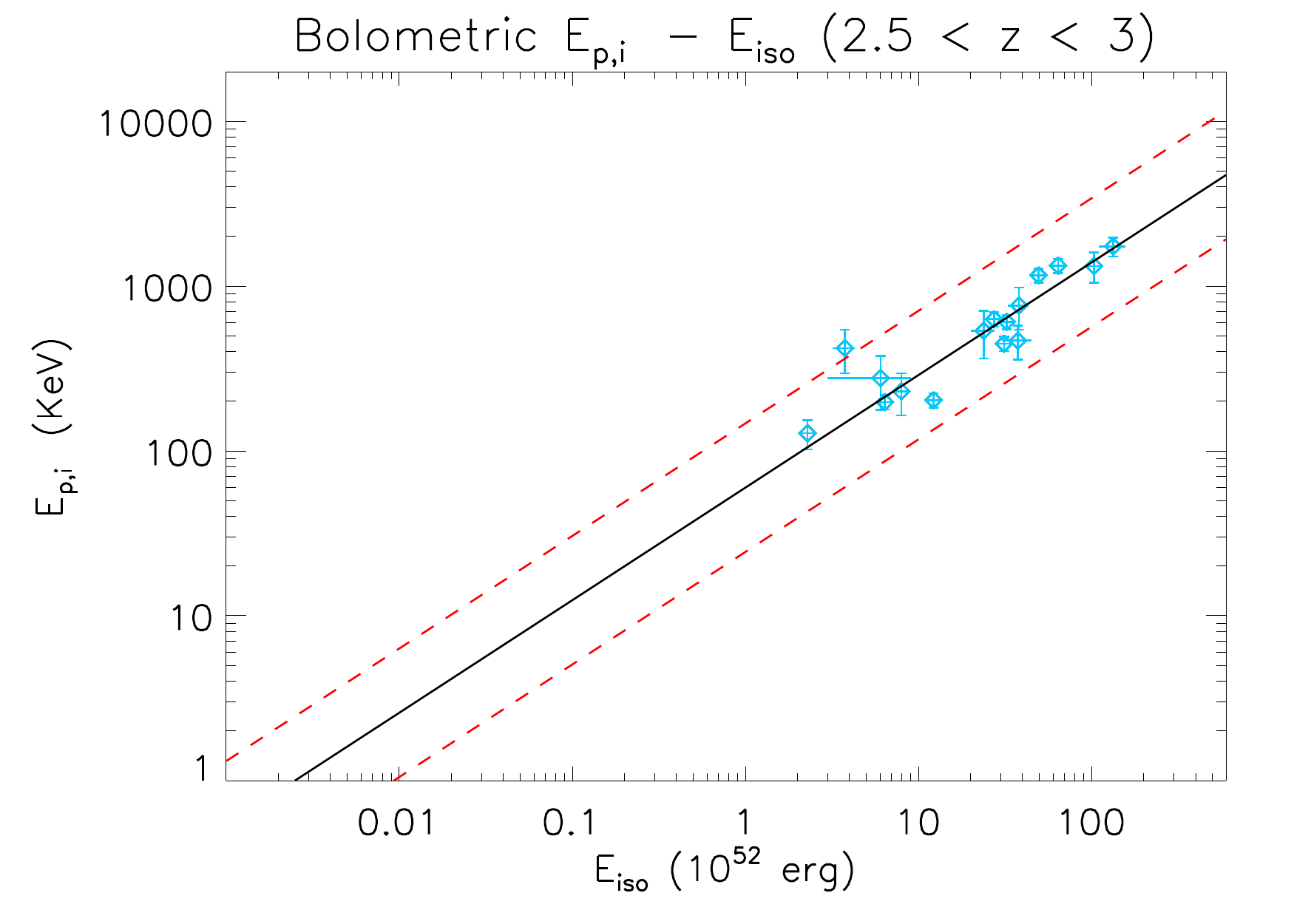}
  \end{minipage}
  \begin{minipage}[b]{0.5\linewidth}
   \includegraphics[width=.85\linewidth]{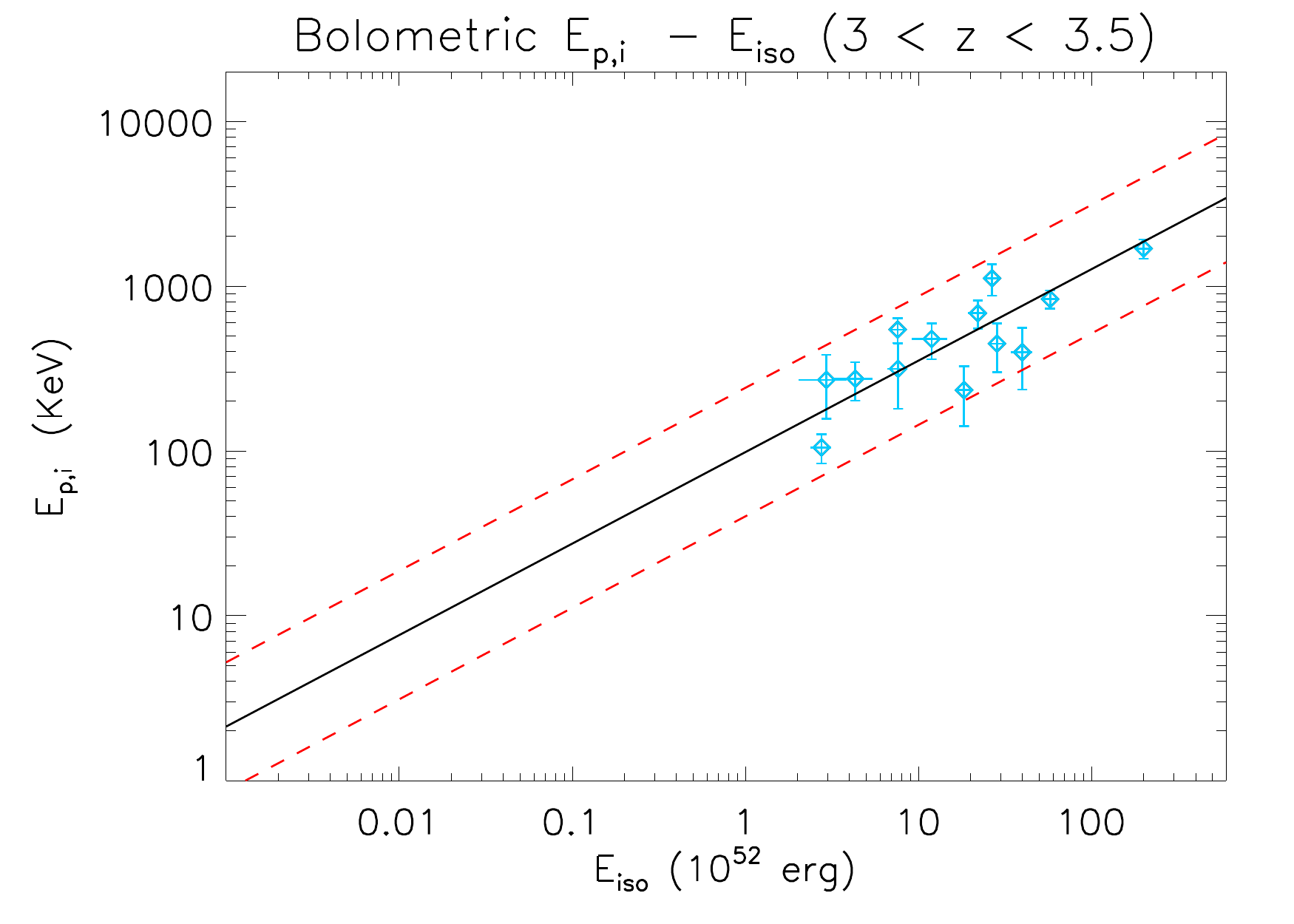}
  \end{minipage}
  \begin{minipage}[b]{0.5\linewidth}
   \includegraphics[width=.85\linewidth]{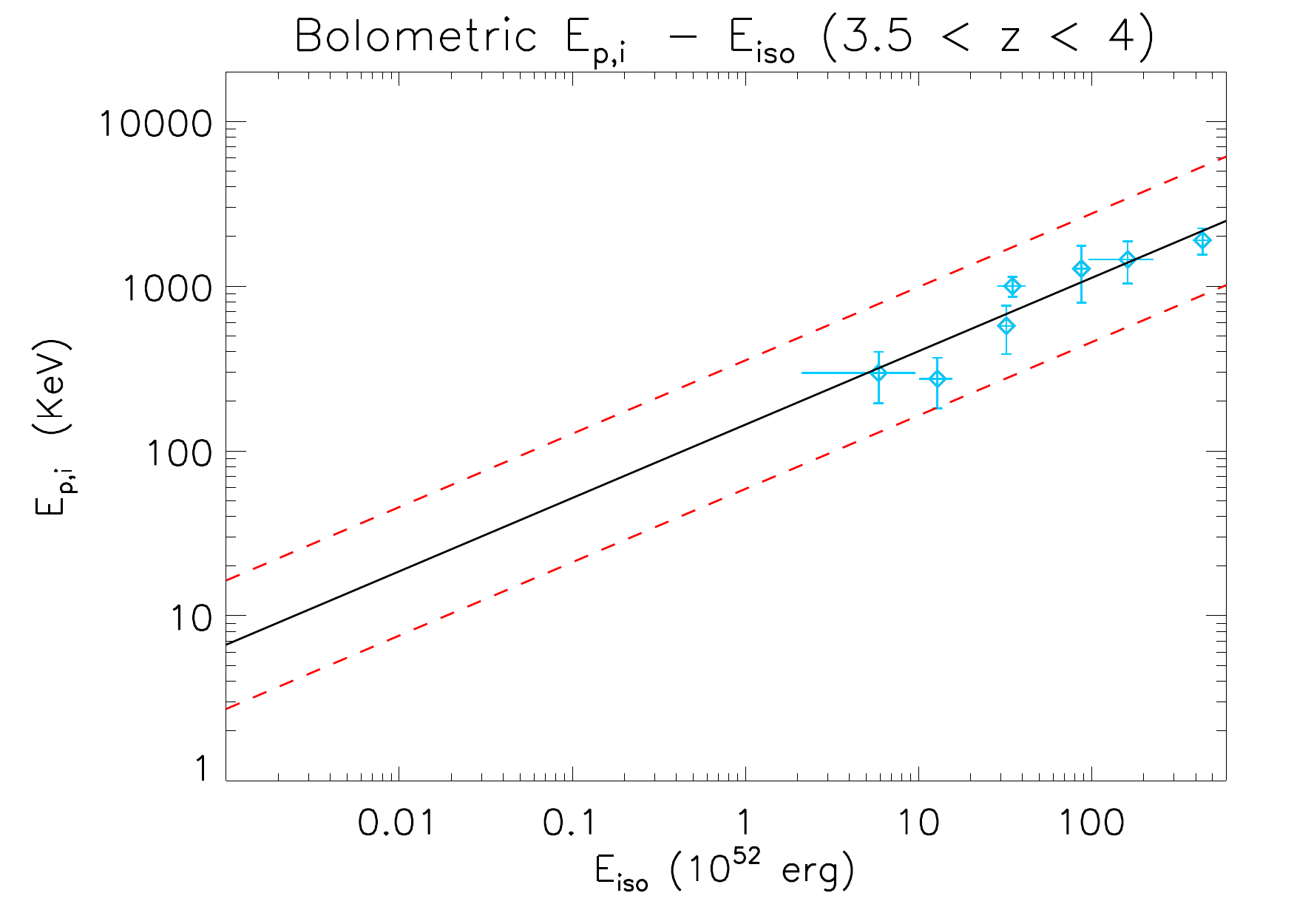}
  \end{minipage}
  \begin{minipage}[b]{0.5\linewidth}
   \includegraphics[width=.85\linewidth]{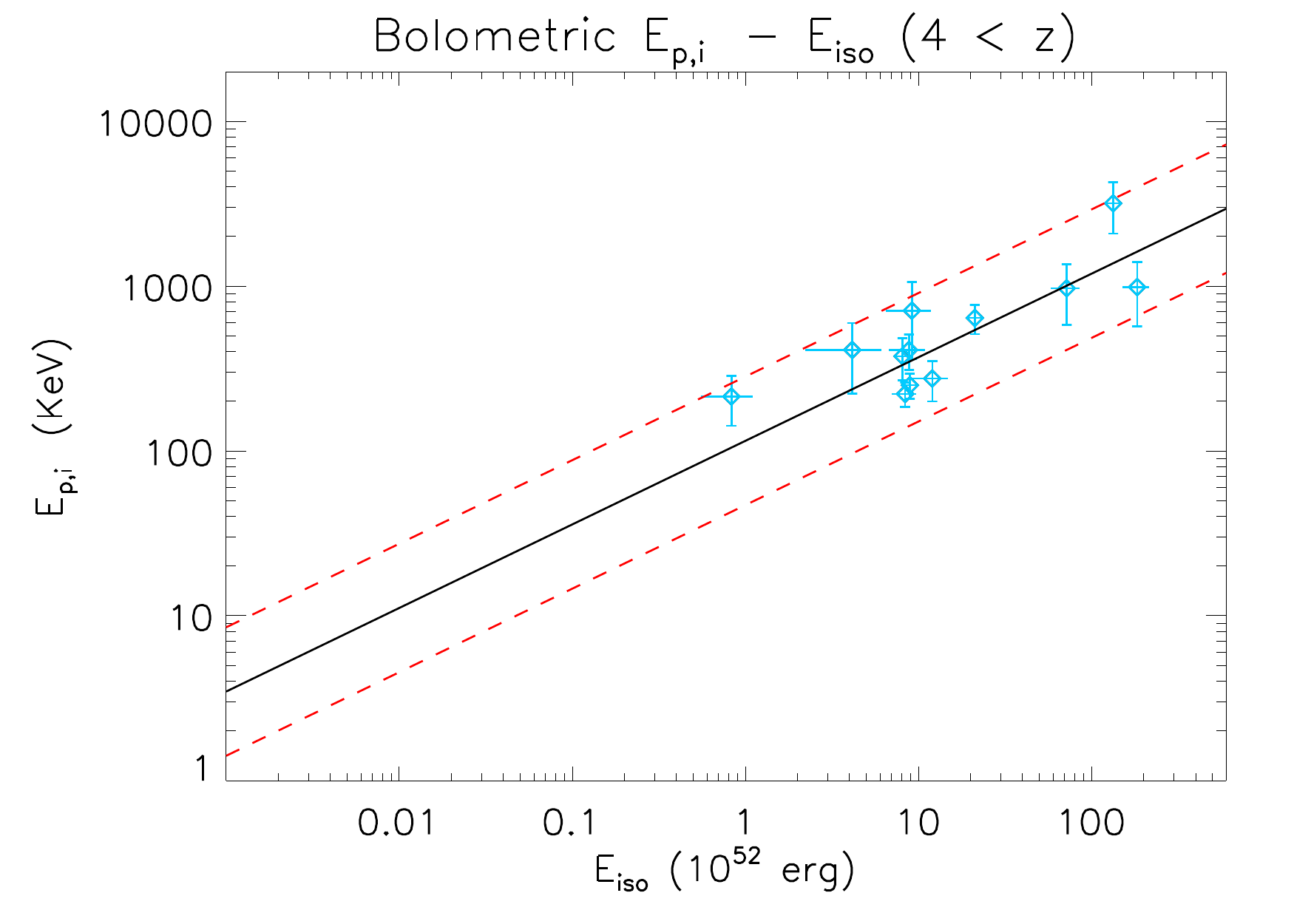}
  \end{minipage}
  \caption{Dependence of time integrated bolometric E$ _{p,i}$$-$E$ _{iso}$ correlation on differeninst redshift b.}
  \end{figure}

\begin{figure}
  \hspace{-0.5cm}
  \begin{minipage}[b]{0.5\linewidth}
   \includegraphics[width=.85\linewidth]{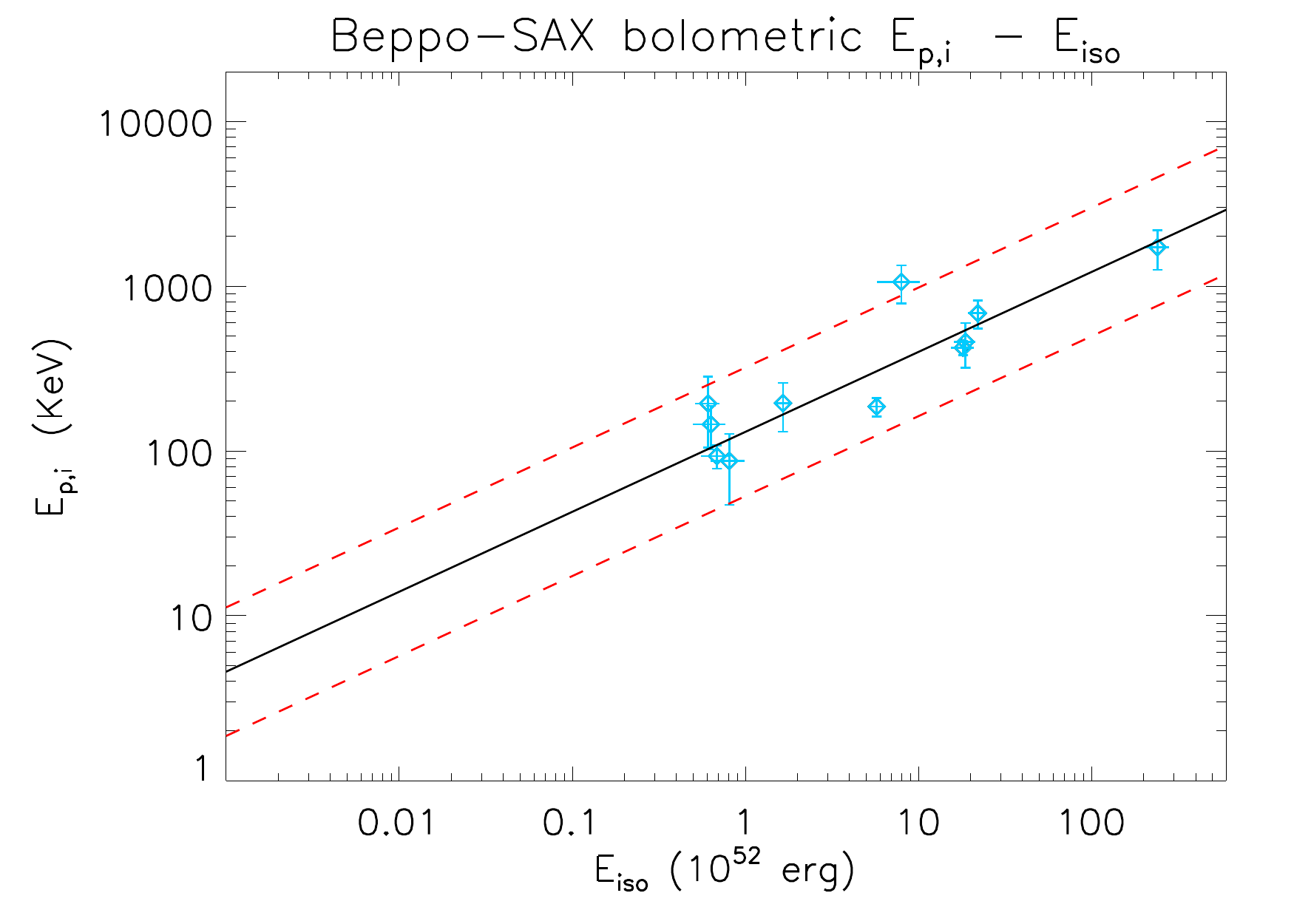}
  \end{minipage}
    \begin{minipage}[b]{0.5\linewidth}
    \includegraphics[width=.85\linewidth]{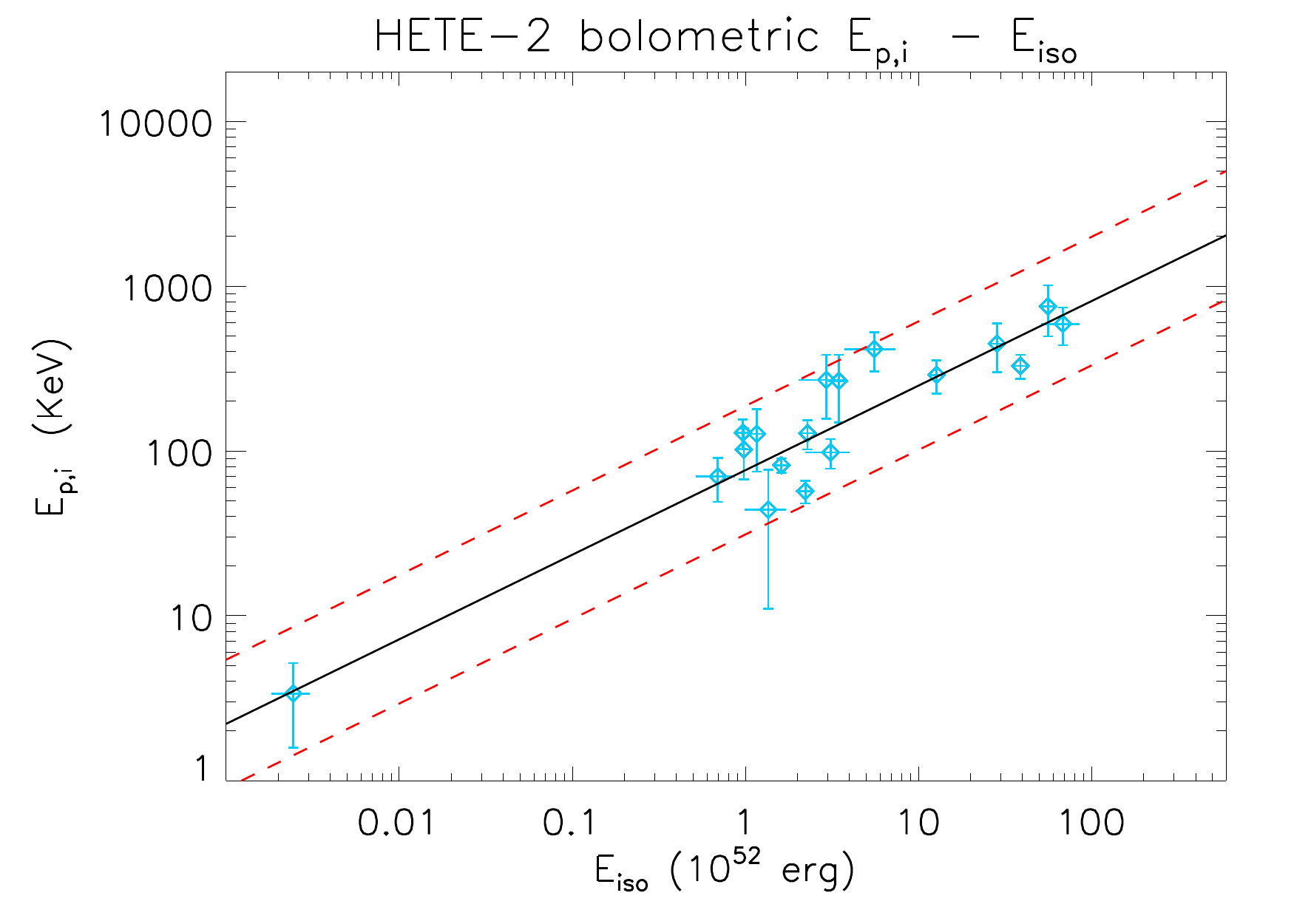}
  \end{minipage}
  \begin{minipage}[b]{0.5\linewidth}
     \includegraphics[width=.85\linewidth]{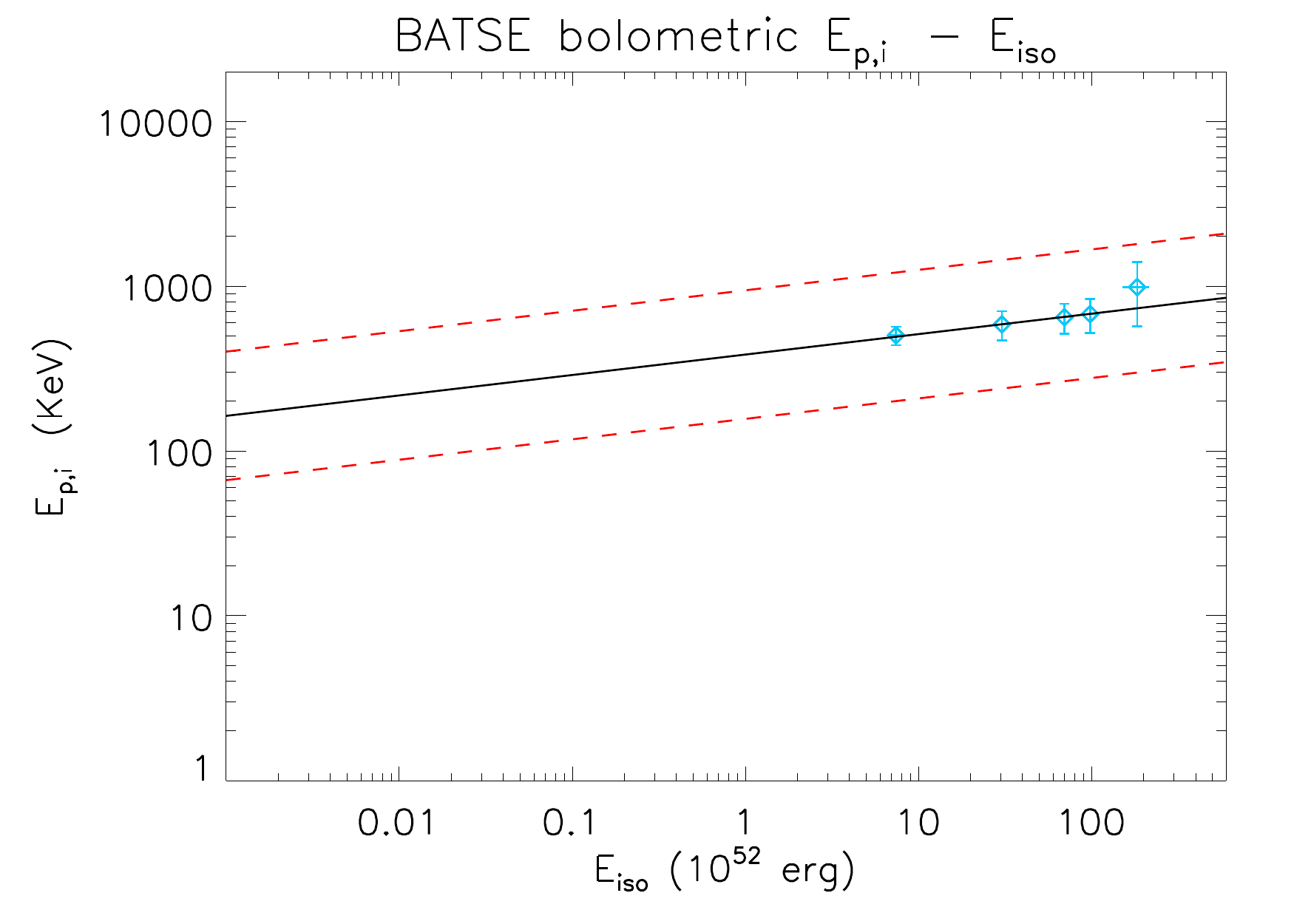}
\end{minipage}
  \begin{minipage}[b]{0.5\linewidth}
   \includegraphics[width=.85\linewidth]{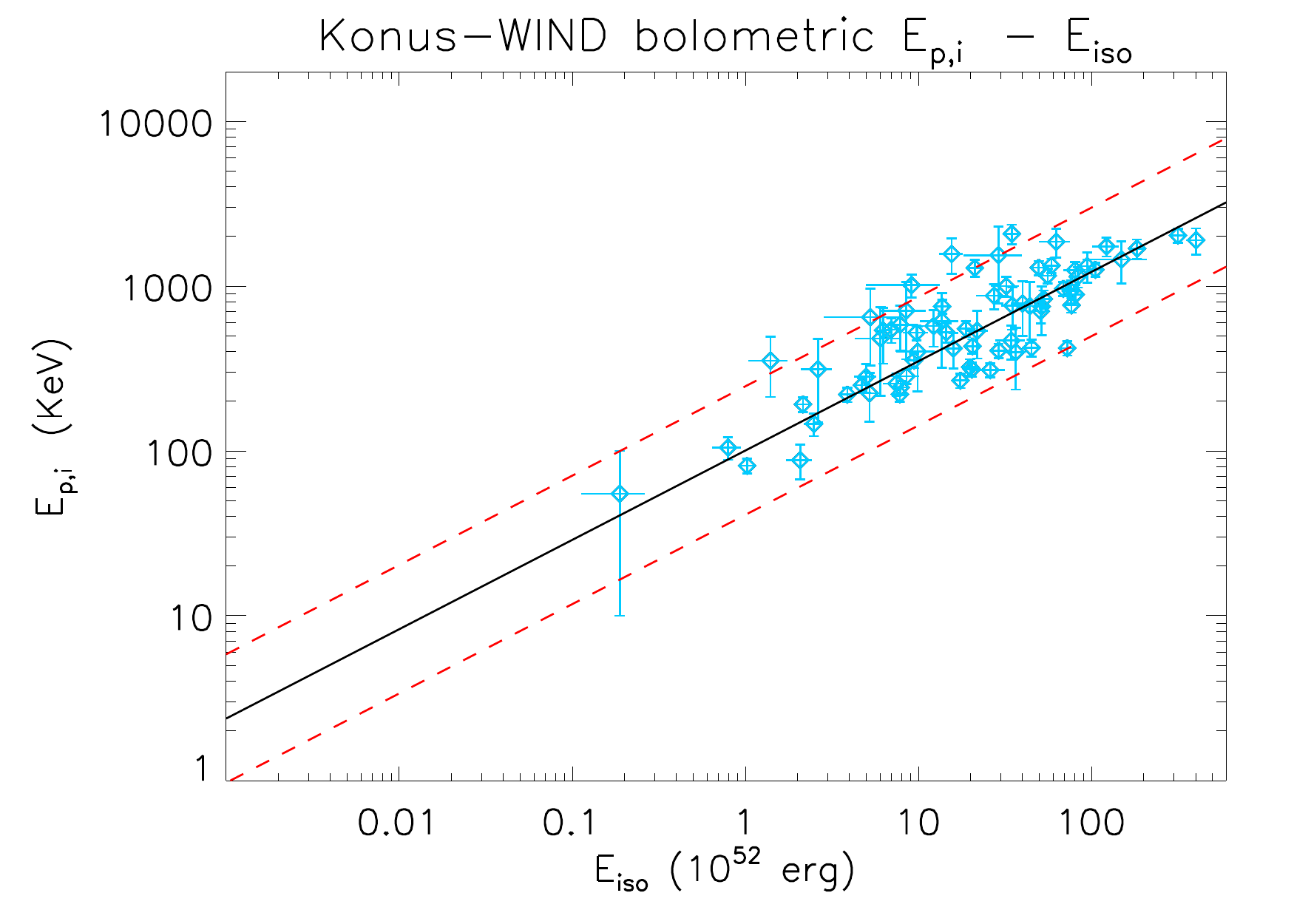}
  \end{minipage}
  \begin{minipage}[b]{0.5\linewidth}
   \includegraphics[width=.85\linewidth]{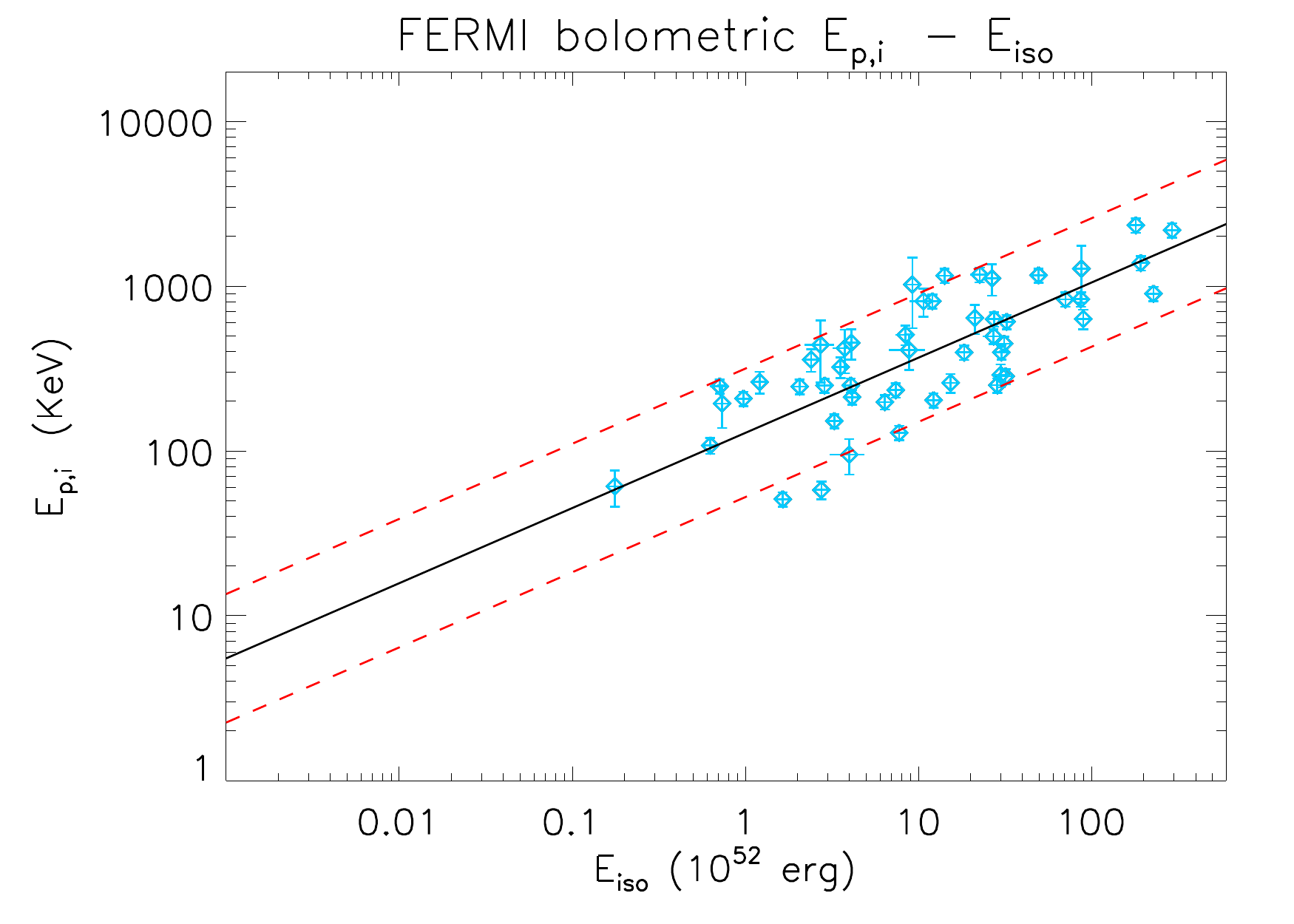}
  \end{minipage}
 \begin{minipage}[b]{0.5\linewidth}
  \includegraphics[width=.85\linewidth]{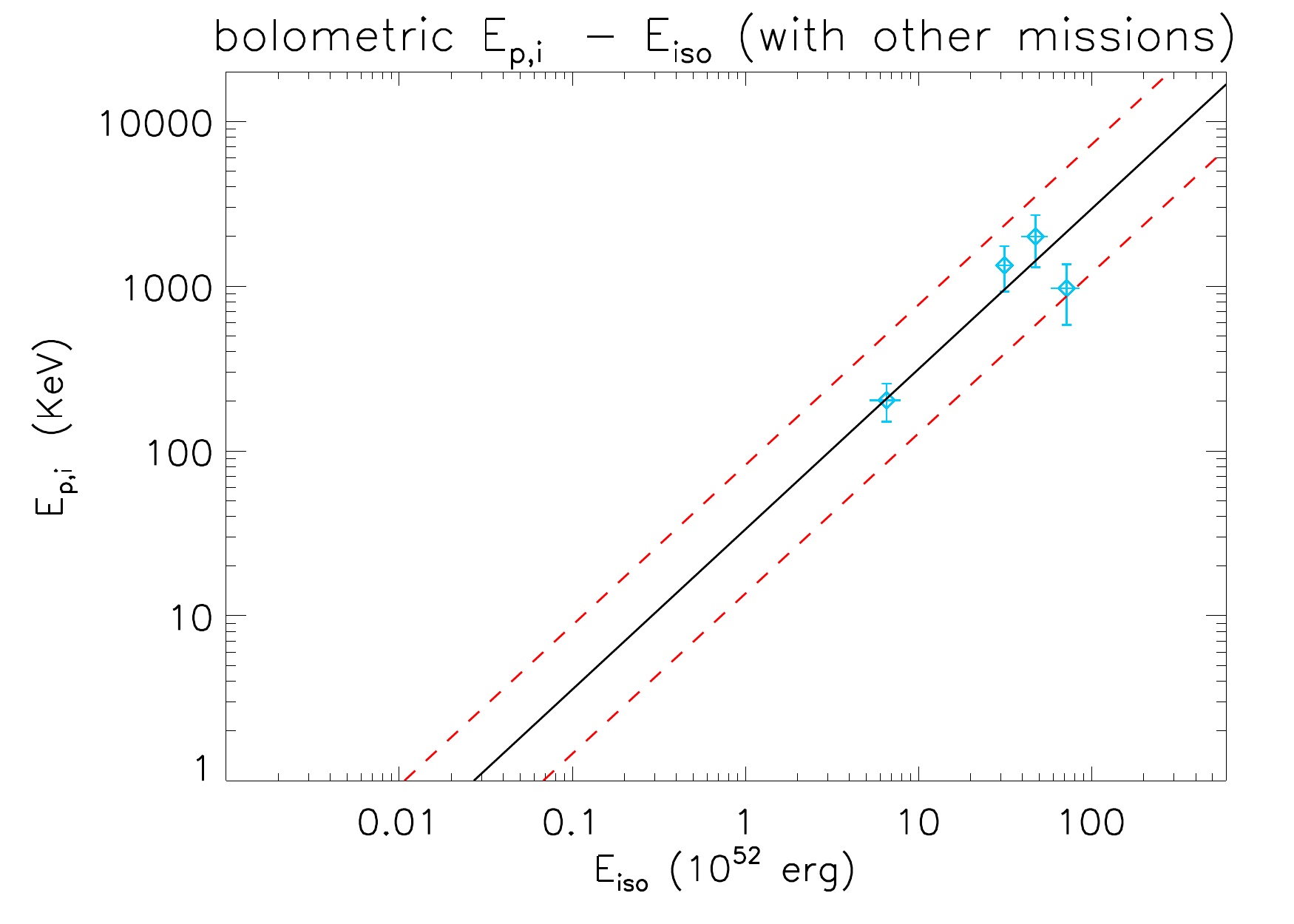}
 \end{minipage}
    \caption{Dependence of the E$ _{p,i}$$-$E$ _{iso}$ correlation on different energy ranges (based on various satellite missions).}
  \label{Ep-Eisoenergy}
  \end{figure}

Furthermore, the Swift satellite, thanks to its capability of providing quick and accurate localization of GRBs,
 thus reducing the selection effects in the observational chain leading to the estimate of GRB redshift, has
 further confirmed the reliability of the $E_{\rm p ,i}$ -- $E_{\rm iso}$ correlation \citep[][]{Amati09,Ghirlanda10,amatidichiara,rmartone,muccino19}. If one divides the GRB sample  on the basis of different high energy satellite missions,  it turns out that the correlation always remains
 within the  same E$ _{p,i}$$-$E$ _{iso}$ fit parameters range (slope $\sim$ 0.5),  as it can be seen from Table \ref{tabdishatesis1}, and Fig.(\ref{Ep-Eisoenergy}).

\begin{table}
\centering
\resizebox{11 cm}{!}{\begin{minipage}{\textwidth}
\begin{tabular}{ cccccc}
\, & \multicolumn{1}{c}{\bf Dependence on energy bins}   \\
\, & \, & \, & \, & \, &  \\
\hline
\multicolumn{1}{c}{\textbf{E$ _{p,i}$$-$E$ _{iso}$ Correlation}} & {\textbf{total GRBs}} & {\textbf{normalization}} & {\textbf{slope}} & {\textbf{scatter}} \\
 \hline
\hline\\
BeppoSAX (2--700 keV)& 11 & 2.11$\pm$0.08 & 0.48$\pm$0.08 & 0.138$\pm$0.057\\
HETE--2 (2--500 keV)& 18 & 1.88$\pm$0.06 & 0.51$\pm$0.06 & 0.133$\pm$0.041\\
Konus--WIND (20 keV--10 MeV)& 72 & 2.03$\pm$0.06 & 0.54$\pm$0.04 & 0.168$\pm$0.019\\
SWIFT (15--150 keV)& 32 & 1.96$\pm$0.03 & 0.55$\pm$0.04 & 0.113$\pm$0.031\\
FERMI (10--30 MeV)& 51 & 2.11$\pm$0.06 & 0.45$\pm$0.05 & 0.236$\pm$0.026\\
\hline
\end{tabular}
\end{minipage}}
\caption{Dependence of the E$ _{p,i}$$-$E$ _{iso}$
correlation on different energy bins (based on
various satellite missions).}
 \label{tabdishatesis1}
\end{table}

  \begin{figure}
 \begin{minipage}[b]{0.9\linewidth}
   \includegraphics[width=.95\linewidth]{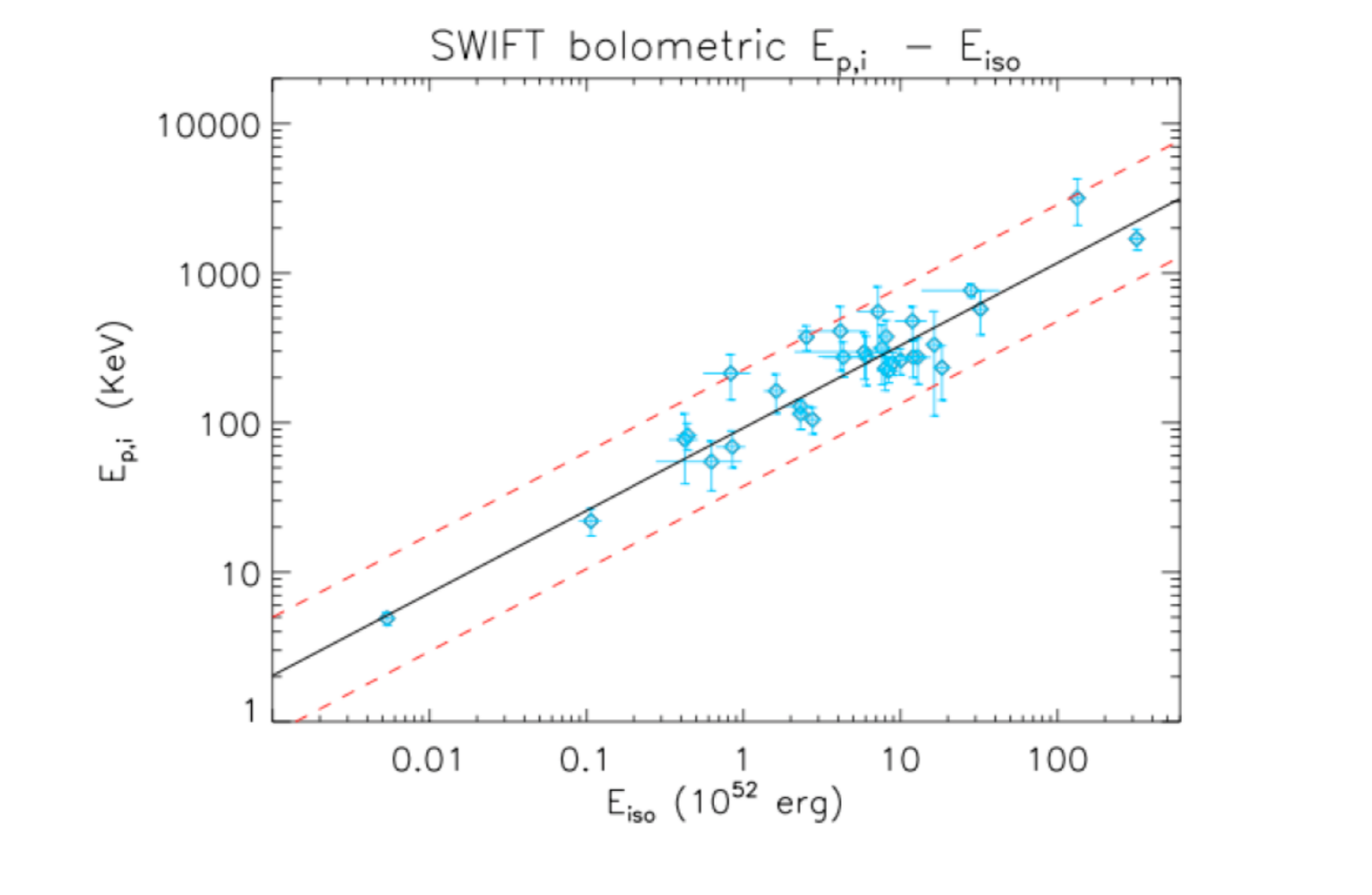}
 \end{minipage}
 \caption{Monochromatic E$ _{p,i}$$-$E$ _{iso}$ correlation from the Swift mission}

\end{figure}

Moreover, based on time-resolved analysis of BATSE, BeppoSAX and Fermi GRBs, it was found that the  $E_{\rm p ,i}$ -- $E_{\rm iso}$  correlation also
holds within each single GRB with normalization and slope consistent with those obtained with time-averaged spectra and energetics/luminosity
\citep[][]{basak,frontera13,Lu12}, confirming the physical origin of the correlation, and providing clues to its explanation. Therefore, it turns out that, at the present stage,  the fit values of the $E_{\rm p ,i}$ -- $E_{\rm iso}$  correlation parameters are marginally affected by selection and/or evolutionary effects, which are less than the intrinsic dispersion (\citep[][]{A-DV,AmatiDisha,A-D,MGRB1,SawantMG}).
\subsubsection{ Calibrating the E$ _{p,i}$$-$E$ _{iso}$ correlation and fitting its parameters }
\begin{figure}
\centerline {\includegraphics[width=8 cm,
height=7 cm]{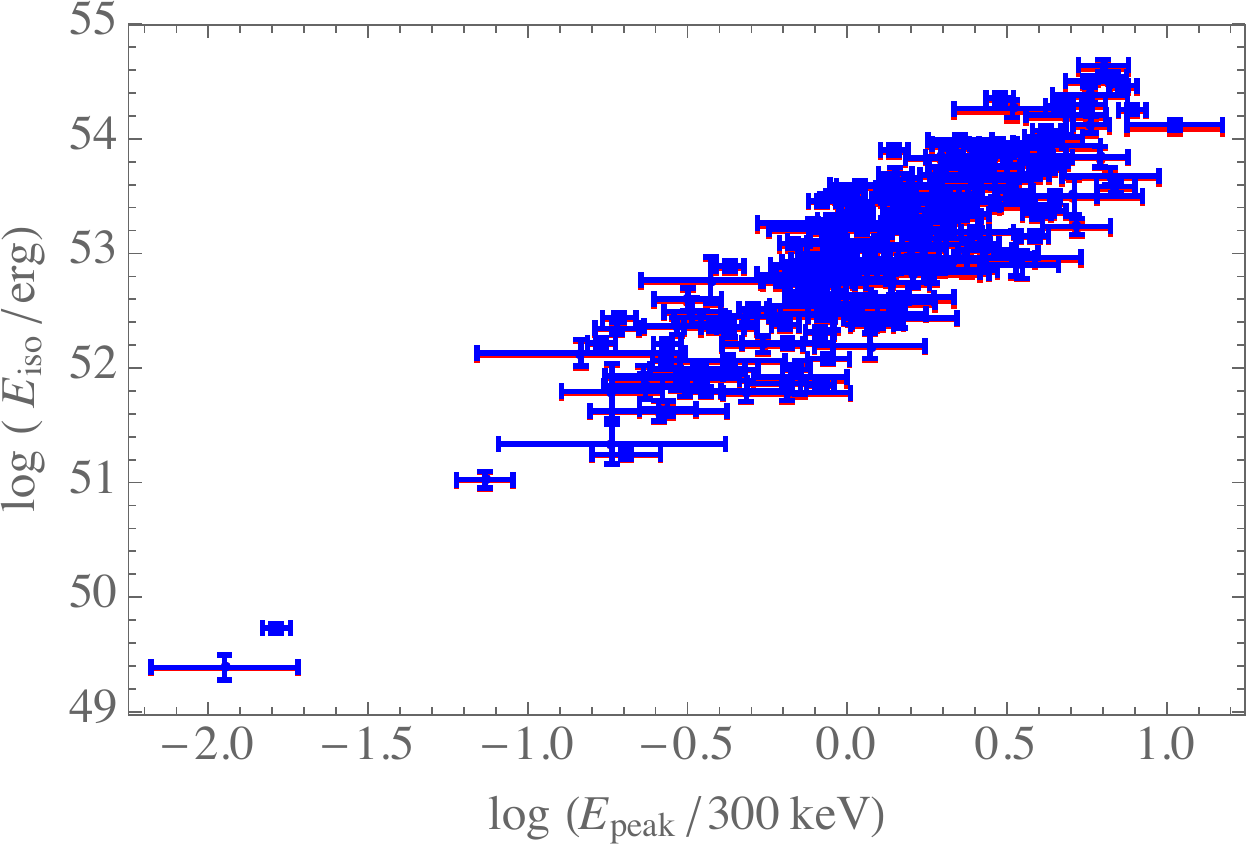}} \caption{
De-evolved (red points) and evolved/original  E$ _{p,i}$$-$E$
_{iso}$ correlation (blue points); there is no noticable
evolution. } \label{Ep_Eisodecon}
\end{figure}

 Here we update  a  local regression
technique, inspired by the standardization of the SNIa with Cepheid variables, and based on Union 2.1 SNIa sample, which we adopted in previous works \citep[][]{MGRB1},\citep[][]{MGRB2b},\citep[][]{Megrb11}. We actually consider a sort of {\it extended}  E$ _{p,i}$$-$E$ _{iso}$ correlation, introducing
 in the calibration terms representing the $z$-evolution, which
are  assumed to be  power-law functions: $g_{iso}(z)=\left(1+z\right)^{k_{iso}}$ and
$g_{p}(z)=\left(1+z\right)^{k_{p}}$, so that $E_{\rm iso}^{'}
=\displaystyle\frac{E_{\rm iso}}{g_{iso}(z)}$ and $E_{\rm p,i}^{'} =\displaystyle\frac{E_{\rm p,i}}{g_{p}(z)}$ are the
de-evolved  quantities \citep[see also ][]{Shirokov20}.Therefore we consider  a correlation with three parameters $a$, $b$, and $k_{iso} - ak_{p}$:
\begin{eqnarray}
 \label{eqamatievol}
&&\log \left[\frac{E_{\rm iso}}{1\;\mathrm{erg}}\right] = b+a \log  \left[
    \frac{E_{\mathrm{p,i}} }{300\;\mathrm{keV}} \right]+\nonumber \\&& + \left(k_{iso} - a k_{p}\right)\log\left(1+z\right)\,.
\end{eqnarray}
We can simplify the redshift dependence term in Eq. (\ref{eqamatievol}), introducing a single average coefficient $\gamma$:
\begin{eqnarray}
 \label{eqamatievol2}
&&\log \left[\frac{E_{\rm iso}}{1\;\mathrm{erg}}\right] = b+a \log  \left[
    \frac{E_{\mathrm{p,i}} }{300\;\mathrm{keV}} \right]+\nonumber \\&& + \gamma\log\left(1+z\right)\,.
\end{eqnarray}
Calibrating this 3D  relation means determining the coefficients
$a$, $ b$, and $\gamma$ and the intrinsic scatter
$\sigma_{int}$. {It is worth noting that}low values of $\gamma$ would
indicate negligible evolutionary effects. In order to calibrate
our de-evolved  relation we  consider a 3D Reichart
likelihood:
\begin{eqnarray}
 \label{reich3dl}
&&L^{3D}_{Reichart}(a,  \gamma, b,  \sigma_{int}) =  \frac{1}{2} \frac{\sum{\log{(\sigma_{int}^2 + \sigma_{y_i}^2 + a^2
\sigma_{x_i}^2)}}}{\log{(1+a^2)}}\,+\nonumber \\ &+& \frac{1}{2} \sum{\frac{(y_i - a x_i -\gamma z_i-b)^2}{\sigma_{int}^2 + \sigma_{x_i}^2 + a^2
\sigma_{x_i}^2}}\,.
\end{eqnarray}
We maximized our likelihood with respect to
$a$ and $\gamma$ since $b$ can be evaluated
analytically by solving the equation
\begin{equation}
{\frac{\partial }{\partial b}L^{3D}_{Reichart}(a, k_{iso},
\alpha, b, \sigma_{int})=0\,,}
\end{equation}
we obtain
\begin{equation}
b = \left [ \sum{\frac{y_i - a x_i-\gamma z_i}{\sigma_{int}^2 + \sigma_{y_i}^2
+ a^2 \sigma_{x_i}^2}} \right ] \left [\sum{\frac{1}{\sigma_{int}^2 + \sigma_{{y_i}}^2 + a^2 \sigma_{x_i}^2}} \right ]^{-1}\,. \label{eq:calca}
\end{equation}
We also used the MCMC method to maximize the likelihood and ran
five parallel chains and the Gelman-Rubin convergence test.
We obtain $a=1.92\pm 0.09$\,, $b=52.7^{+0.04}_{-0.03}$\,, $\sigma_{\rm int}=0.35_{-0.05}^{+0.04}$\,, $\gamma=-0.07\pm 0.14$, thus confirming that the evolutionary effects, not included in the intrinsic dispersion, can be, at this stage, neglected, as shown in Fig. (\ref{Ep_Eisodecon}).

\subsubsection{Further investigations on the calibration of the $E_{\rm p,i}$ -- $E_{\rm iso}$ correlation}
\begin{figure}
\centerline{ \includegraphics[width=.9\linewidth,height=.8\linewidth]{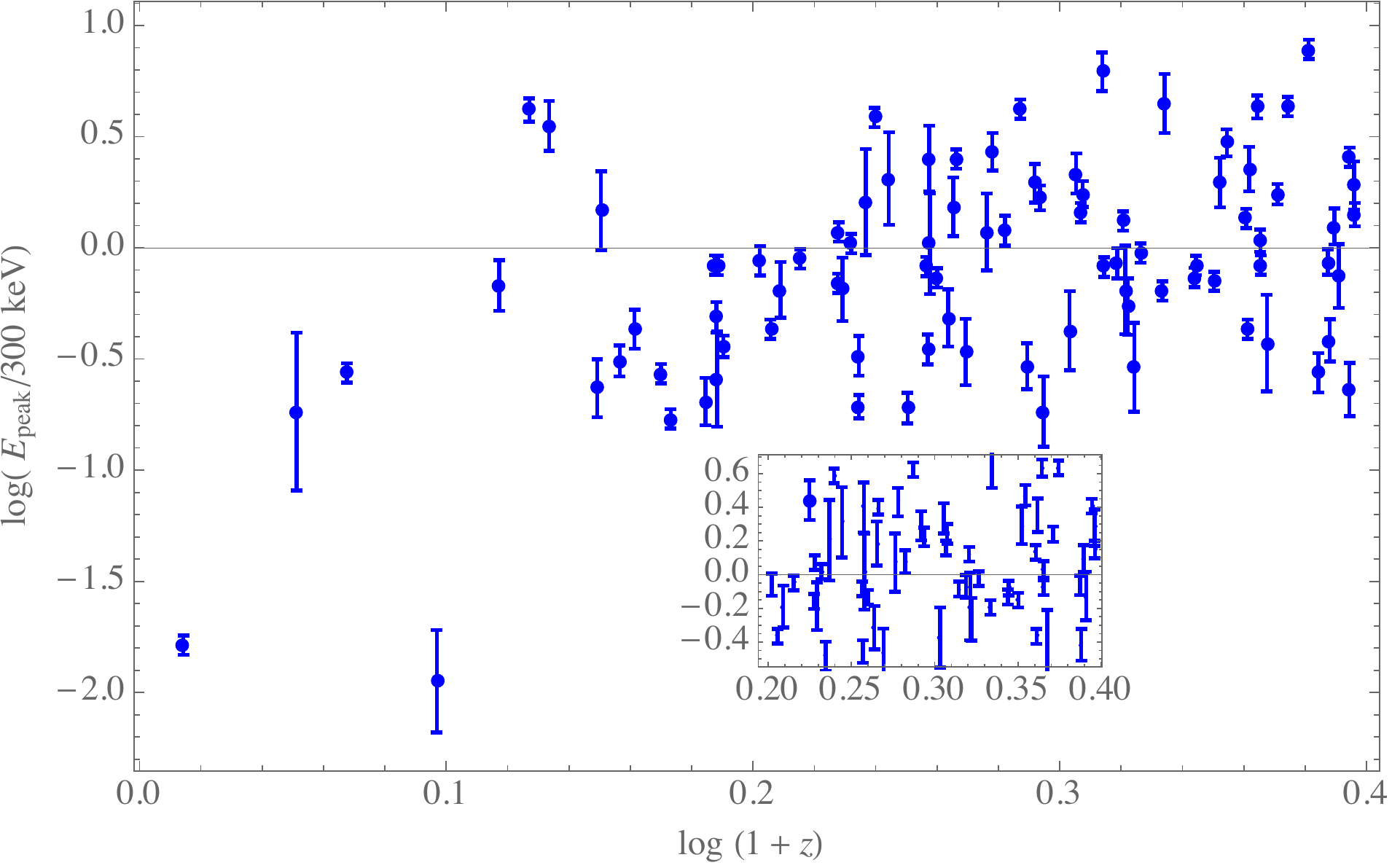}}
  \caption{ Redshift  distribution of $\log E_{\rm p,i} /(300 keV)$ for the subsample used for our calibration. }
\label{selection1}
\end{figure}

\begin{figure}
\centerline{ \includegraphics[width=.9\linewidth,height=.8\linewidth]{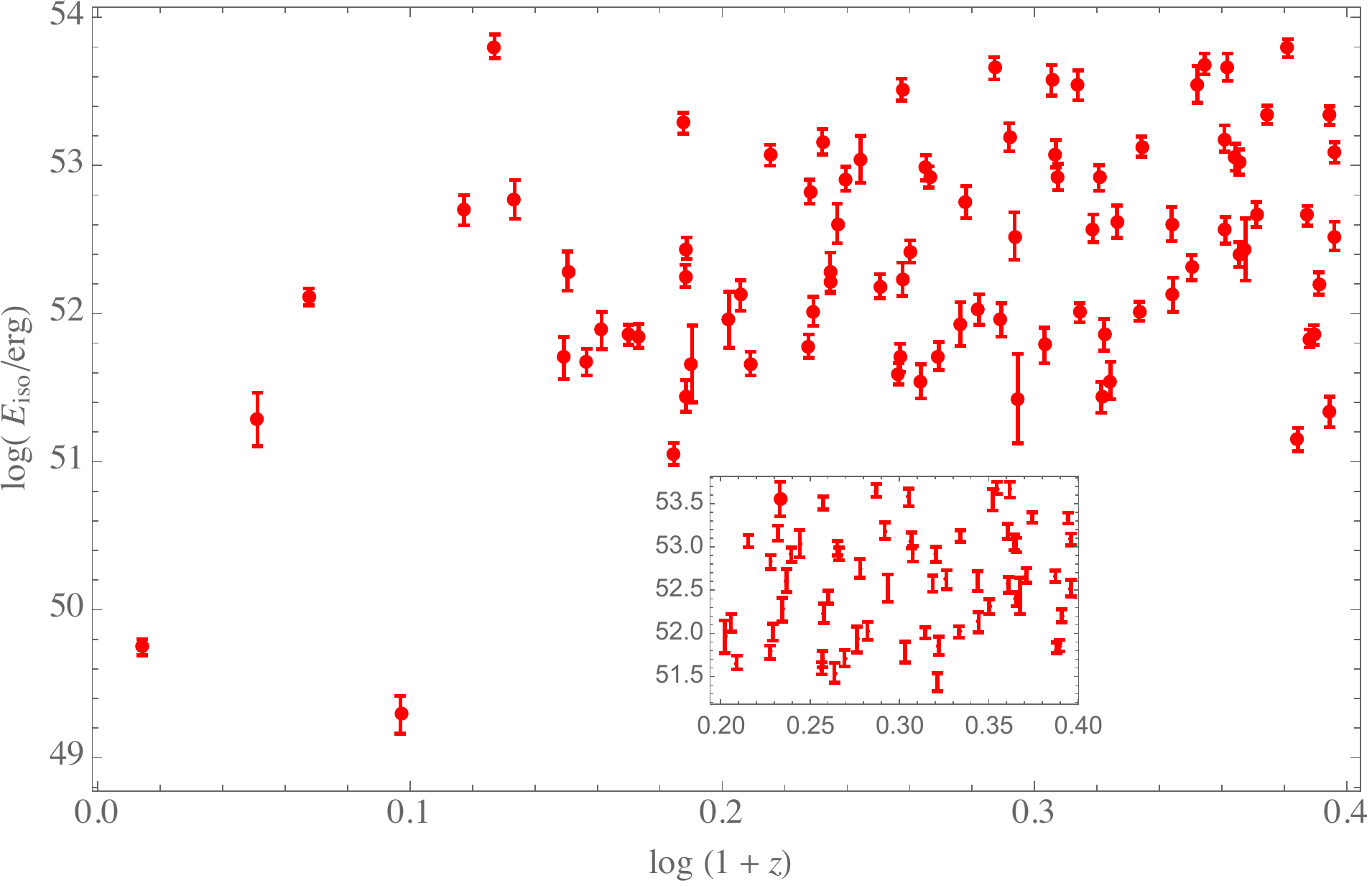}}
\caption{Redshift  distribution of $\log E_{\rm iso} /(erg)$ for the subsample used for our calibration. }
\label{selection2}
\end{figure}

In this section we discuss some aspects related to the calibration of the  $E_{\rm p,i}$ -- $E_{\rm iso}$ relation, and its impact on reliability of the GRBs as distance indicators. Actually, we apply some  filters on $ E_{\rm p,i}$ and $E_{\rm iso}$,  reducing  the sample to 60 objects, as shown in Figs. (\ref{selection1} and \ref{selection2}), where the $E_{\rm p,i}$ and $E_{\rm iso}$ are {\it homogeneously} distributed in the sample. Note that the measured values of  $E_{\rm p,i}$ and  $E_{\rm iso}$ are not systematically larger at lower redshift than at higher redshifts.

\begin{figure}
\centerline{ \includegraphics[width=.9\linewidth,height=.5\linewidth]{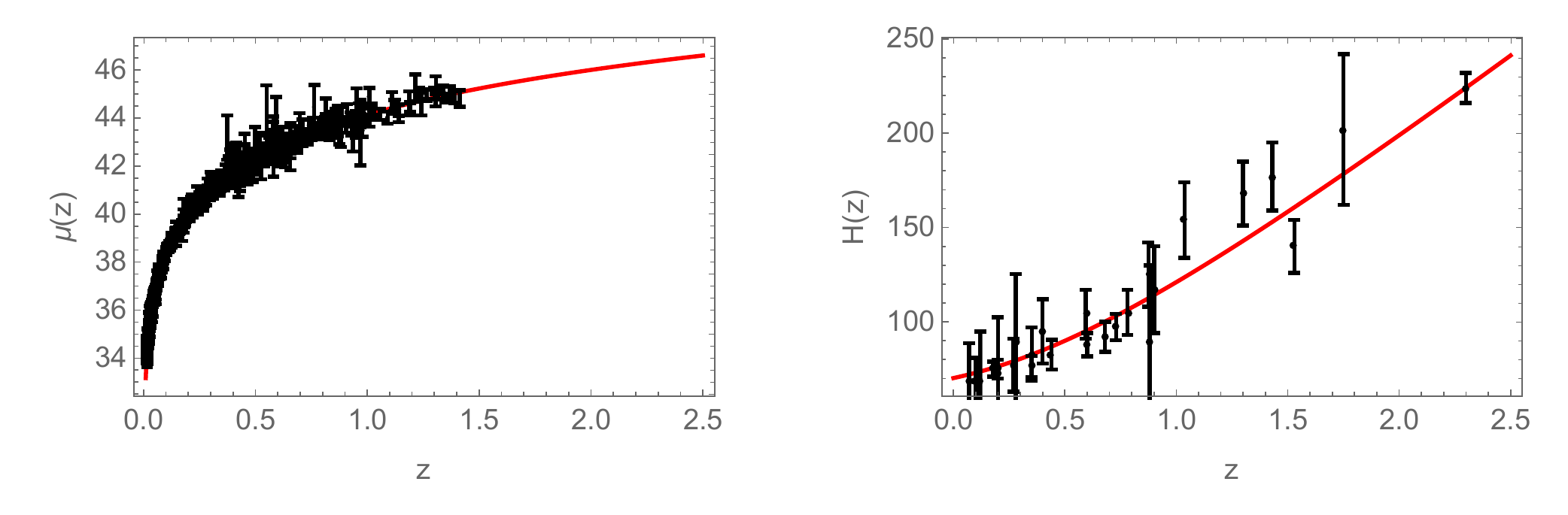}}
\caption{Comparison between the observational data and theoretical predictions for the distance modulus (left panel), and the Hubble function, $H(z)$, (right panel).}
\label{figapp}
\end{figure}

From the Reichart likelihood analysis we find that  $a =2.0_{-0.13}^{+0.14}$,  $b=52.5^{+0.04}_{-0.03}$\,,$\sigma_{\rm int}=0.38\pm 0.06$\,, $\gamma=0.02\pm 0.09$.
The fit  values of the correlation parameters are fully compatible (at 3 $ \sigma$) with the results obtained from the full dataset. Moreover it is worthwhile to note that we have been forced to limit our analysis to the redshfit range $0.015 \leq z \leq1.414$, where we apply our calibration technique. In order to extend our analysis to a wide range of redshifts we build up a further calibration procedure  which updates the procedure described in \citep[][]{Megrb11, MGRB2b}, based on an approximate function for the luminosity distance.
This function $d_L^{app}(z)$ has the form
\begin{equation}
d_L^{app}(z)=\frac{c}{H_0}\frac{z (z+1)^2}{\sqrt{d_3 z^3+\left(d_2 z^2+d_1 z+1\right){}^2}}\,,
\label{approxnew}
\end{equation}
where $d_1$, $d_2$ and $d_3$ are constants.  To estimate these parameters we have simultaneously fitted the SNIa samples and the $H(z)$ measurements. Actually it turns out that we can determine an approximate function for the $H(z)$, from $d_L^{app}(z)$, according to the relation:
\begin{equation}
\frac{1}{H^{app}(z)}=\frac{d}{d z}\left(\frac{d_L^{app}(z)}{1+z}\right)\,.
\label{happ}
\end{equation}
We find that $d_1=1.18\pm 0.03$, $d_2=0.30\pm 0.04$, $d_3=0.49\pm 0.1$. In Fig. (\ref{figapp}), we plot the data with the best fit $d_L^{app}$ and $H^{app}(z)$ respectively.

Therefore we can use the approximate function $d_L^{app}(z)$ to calibrate the $E_{\rm p,i}$ -- $E_{\rm iso}$ correlation, following the procedure described above. Using all the data, we find that $a=1.84\pm 0.08$, $b=52.7\pm 0.05$\,,$\sigma_{\rm int}=0.37\pm 0.04$\,, $\gamma=0.03\pm 0.1$.
As the last check we applied the same filters on $ E_{\rm p,i}$ and $E_{\rm iso}$  as before, selecting a sub-sample of 128 GRBs, and calibrated the  $ E_{\rm p,i}$ -- $E_{\rm iso}$ correlation.
We find that $a=1.94\pm 0.09$, $b=52.4\pm 0.04$\,, $\sigma_{\rm int}=0.38\pm 0.05$\,, $\gamma=0.02\pm 0.15$. Also in this case it turns out that  the calibration parameters $a$,  $\gamma$ and $\sigma_{int}$ are fully consistent with our SNIa-calibration technique, and that all the possible systematics and evolutionary effects are within the intrinsic dispersion $\sigma_{int}$, as already discussed in literature ( see, for instance, \citep[][]{Amati08, A-DV, amatidichiara}).
However it is worth noting that when future GRB missions will
substantially increase the number of GRBs usable to construct
the $E_{\rm p,i}$ -- $E_{\rm iso}$  correlation  up to redshift
$ z \simeq 10$ , they may shed new light on the properties of this important correlation.
\subsubsection{Bilding up the Hubble diagram}
\begin{figure}
\centerline{\includegraphics[width=.9\linewidth,height=.8\linewidth]{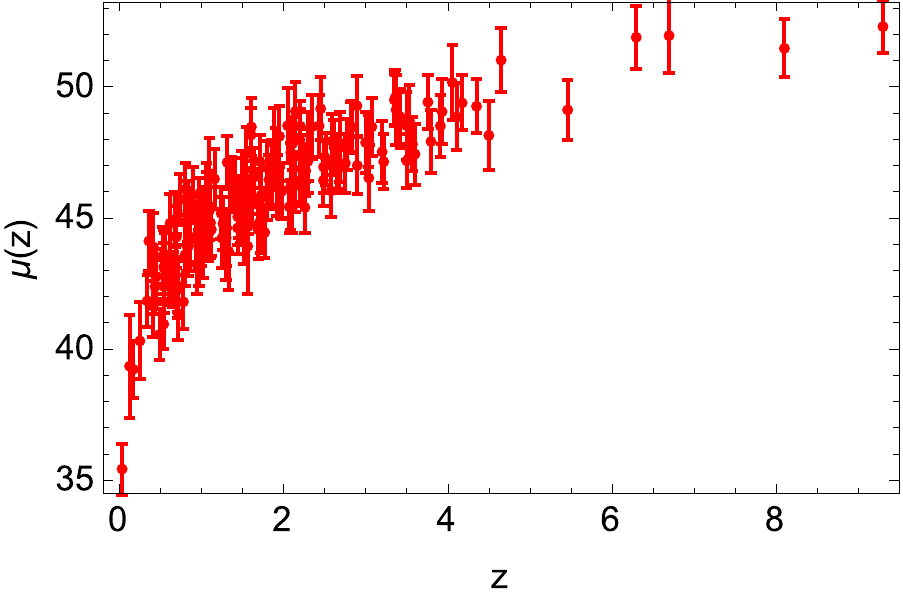}}
\caption{Distance modulus $\mu(z)$ for the calibrated  GRB Hubble diagram obtained by fitting the   $E_{\rm p,i}$ -- $E_{\rm iso}$ relation.}
\label{GRBHDe-evolved}
\end{figure}

\begin{figure}
\centerline{ \includegraphics[width=.9\linewidth,height=.8\linewidth]{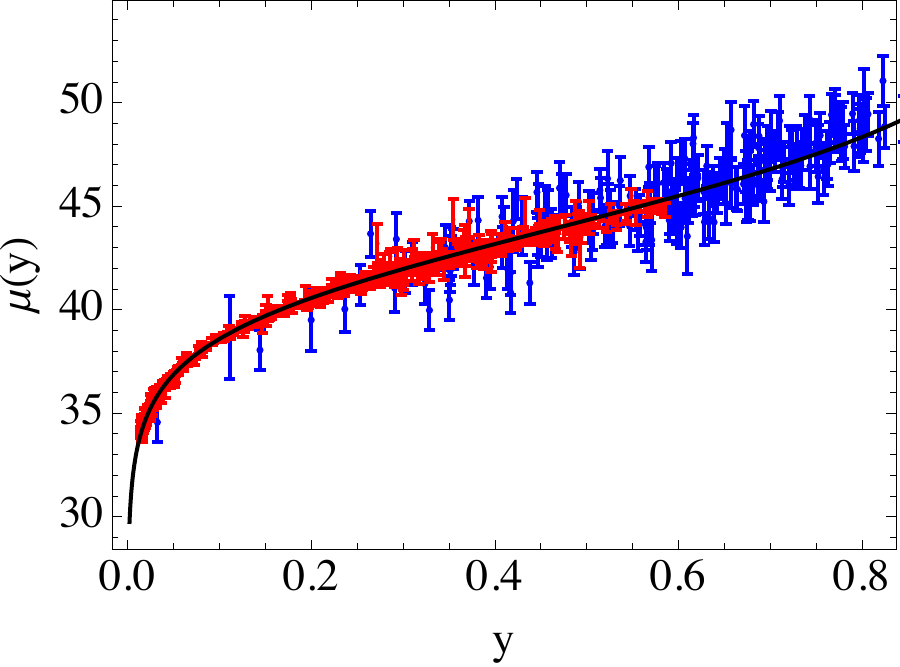}}
\caption{GRB (blue points) and SNIa (red points) superposed Hubble diagrams after $\mu_0$ has been evaluated for the best fit CPL model (see the section below). Here  $\displaystyle y=\frac{z}{1+z}$.}
\label{GRBHDsnmatch}
\end{figure}

After fitting the  correlation
and estimating its parameters, we used them to construct the GRB
Hubble diagram. We recall that the luminosity distance of a GRB
with redshift $z$ is
\begin{equation}\label{lumdist}
d_L(z) = \left( \frac{E_{\rm iso}(1 + z)}{4 \pi  S_{bolo}}\right)^{1/2}.
\end{equation}
The uncertainty of $d_L(z)$ was estimated through the
propagation of the measurement errors of the pertinent
quantities. In particular, recalling that our correlation
relation  can be written as a linear relation, as in  Eqs.
(\ref{eqamatievol}, \ref{eqamatievol2}), the error on the distance dependent
quantity $y=\log \left[\frac{E_{\rm
iso}}{1\;\mathrm{erg}}\right]$ was estimated as
\begin{equation}
\sigma(y) = \sqrt{a^2 \sigma^2_{x}+\sigma_{a}^2 x^2 +\sigma_{b}^2+ \sigma_{int}^2},
\label{eq:siglogy}
\end{equation}
where $x= \log  \left[\frac{E_{\mathrm{p,i}} }{300\;\mathrm{keV}} \right]$\,, $\sigma_b$
is properly evaluated through the Eq. (\ref{eq:calca}), which implicitly  defines b as
a function of $a$ and $\sigma_{int}$,
and is then added in quadrature to the uncertainties of the other terms entering
Eq.(\ref{lumdist}) to obtain the total uncertainty. It turns out that
\begin{eqnarray}
&& 5 \log{d_L(z)} = \left(\frac{5}{2}\right)\left\{b+a\log
\left[\frac{E_{\mathrm{p,i}} }{300\;\mathrm{keV}}\right]+ \right.\nonumber
\\&& +\left. \left(\gamma+1\right) \log\left(1+z\right)-\log\left(4 \pi
S_{bolo}\right)+\mu_0 \right\},
\end{eqnarray}
where $\mu_0$ is  a normalization parameter. Actually the distance modulus of GRBs are not absolute, thus this cross-calibration parameter is needed to match the GRB Hubble diagram and the one of supernovae. If $\mu_0$ cannot be determined,  we can only use the shape of the Hubble Diagram to constrain the cosmological parameters such
as $\Omega_m$\, and $\Omega_{\Lambda}$, with no information on
$H_0$, which is degenerate with the $\mu_0$ parameter.  It turns out that $\mu_0$ depends on the fiducial cosmological model and its parameters; however in all models considered in our analysis $\mu_0\simeq 0.4$. In Fig. (\ref{GRBHDe-evolved}) we plot the GRB Hubble diagram, and in Fig. (\ref{GRBHDsnmatch}) we show that this diagram matches the Union 2.1 Hubble diagram, then $\mu_0$ has been evaluated for the CPL best fit model, discussed above. In order to make this comparison clearer, we use the auxiliary variable $\displaystyle y=\frac{z}{1+z}$, which map the $z$-range $\left[0,\infty\right)$ into the $y$-interval $\left[0,1\right]$. It turns out that the GRBs are the natural continuation of SNIa in the Hubble diagram.
The listed data are available on request to the authors.

\subsection{Direct H(z) measurements}
 The direct measurements of Hubble parameters are complementary probes to constrain
the cosmological parameters and investigate the dark energy \citep[][]{Jimenez}. The Hubble parameter, defined as $H(z) =\displaystyle \frac{\dot a}{a}$,  where $a$ is the scale factor,
depends on the differential variation of the cosmic time with redshift as
\begin{equation}
H(z)=- \frac{1}{1+z} \left( \frac{dt}{dz}\right)^{-1}\,.
\label{hz}
\end{equation}
The $\displaystyle \left( \frac{dt}{dz}\right)$
can be measured  using the so-called cosmic chronometers. $dz$ is obtained from spectroscopic
surveys with high resolution, and  the differential evolution of the age of the Universe $dt$ in the  redshift interval
$dz$ can be measured provided that  optimal probes of the aging of the Universe, that is, the cosmic chronometers, are
identified. The most reliable cosmic chronometers observable at high redshift are old early-type galaxies that evolve passively on a
timescale much longer than their age difference, which formed the vast majority of their stars rapidly and  early  and
have not experienced  subsequent major episodes of star formation or merging. Moreover, the Hubble parameter can also
be obtained from the BAO measurements: by observing the typical acoustic scale in the line-of-sight direction, it is
possible to extract the expansion rate of the Universe at a certain redshift. We used a list of $28$ direct $H(z)$
measurements in the redshift range $z\sim 0.07-2.3$ compiled by \citep[][]{farooqb}.

\section{Statistical analysis}\label{StatA}
To test the dark energy models described above, we use a Bayesian approach based on the MCMC method. In order to set
the starting points for our chains, we first performed a preliminary and standard fitting procedure to maximize the
likelihood function ${\cal{L}}({\bf p})$:
\begin{eqnarray}
\footnotesize
{\mathcal{L}}({\bf p}) & \propto & \frac {\exp{(-\chi^2_{SNIa/GRB}/2)}}{(2 \pi)^{\frac{{\cal{N}}_{SNIa/GRB}}{2}} |{\bf C}_{SNIa/GRB}|^{1/2}}  \nonumber\\ ~ & \times & \frac{\exp{(-\chi^2_{H}/2})}{(2 \pi)^{{\cal{N}}_{H}/2} |{\bf C}_{H}|^{1/2}} \,.
\label{defchiall}
\end{eqnarray}
Here
\begin{equation}
\chi^2(\mathrm {\bf p}) = \sum_{i,j=1}^{N} \left( x_i -
x^{th}_i(\bf p)\right)C^{-1}_{ij}  \left( x_j - x^{th}_j(\bf
p)\right) \,, \label{eq:chisq}
\end{equation}
 $\bf p$ is the set of parameters, $N$ is the number of data points,  $\mathrm x_i$ is the $i-th$ measurement;
$ x^{th}_i(\bf p)$ indicate the theoretical predictions for
these measurements that depend on the parameters $\bf p$.
$C_{ij}$ is the covariance matrix (specifically, ${\bf
C}_{SNIa/GRB/H}$ indicates the SNIa/GRBs/H  covariance matrix). Eq. (\ref{defchiall})  includes  $\sigma_{int} $ term to allow for intrinsic scatter in the data sets.

\begin{table}
\begin{center}
\renewcommand{\arraystretch}{1.5}
\begin{tabular}{c c  }

\hline
\hline
Parameters & Priors \\
\hline
$\Omega_m$ & $(0,1)$ \\
$w_0$ & $(-3,3)$\\
$w_1$ & $(-3,3)$\\
$\Omega_e$ & $(0,0.1)$ \\
$\mathcal{ H}_0$  & $(0.5,1.5)$ \\
\hline
\hline
\end{tabular}
\caption{Priors for parameters estimate in the MCMC numerical analysis.}
 \label{tabprior}
\end{center}
\end{table}

In our analysis we include only flat priors on the  typical parameters of the considered cosmological models (see Table {\ref{tabprior}}, with the exception of the Hubble constant, $h$, for which we consider a gaussian prior accounting for the local determination of the Hubble constant  by the SHOES collaboration $(h_S, \sigma_S) = (0.742, 0.036)$ \cite{shoes}.
We actually consider the term\,:

\begin{equation}
{\cal{L}}({\bf h}) = \frac{\exp{\left [ - (h_{S} - h)^2/ 4\sigma_{S}^2 \right ]}}{\sqrt{4 \pi \sigma_{S}^2}}\,.
\label{eq: deflike}
\end{equation}

\begin{figure}
\begin{minipage}[b]{0.9\linewidth}
\centerline{\includegraphics[width=.7\linewidth,height=.6\linewidth]{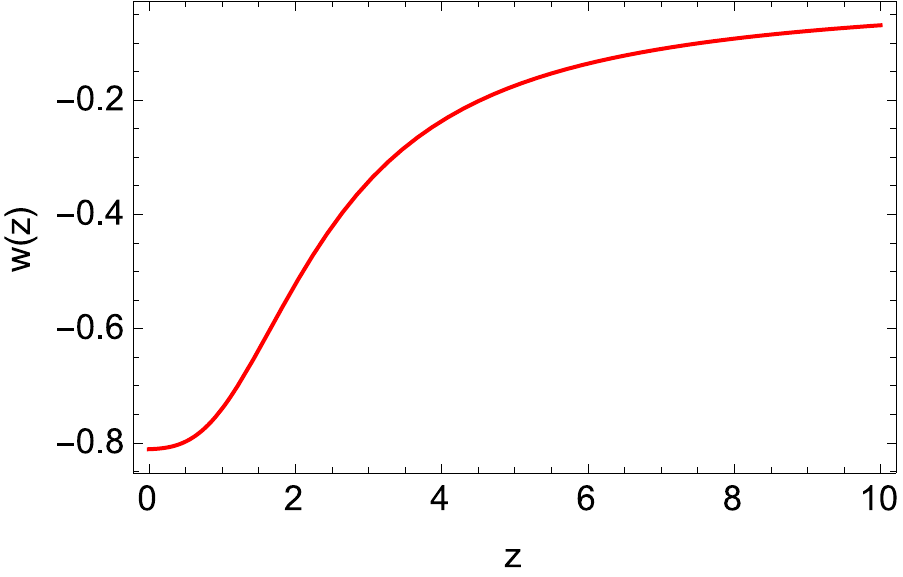}}
 \end{minipage}
\caption{Redshift dependence of the EDE EOS corresponding to the best fit values of the parameters. }
\label{wede}
\end{figure}

\begin{figure}
\centerline{\includegraphics[width=.9\linewidth,height=.8\linewidth]{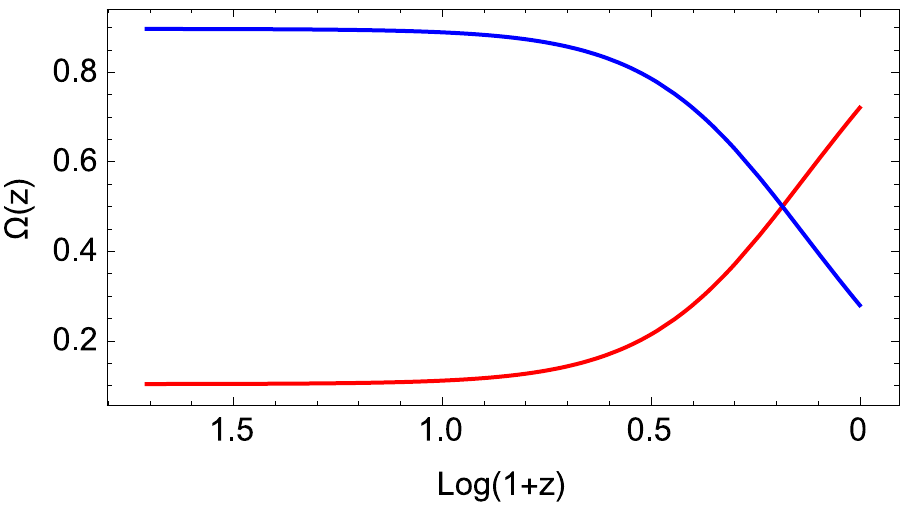}}
 \caption{ Redshift
dependence of the $\Omega$ parameters for the EDE model
corresponding to the best fit values of the parameters:
$\Omega_m(z)$ is shown in blue, and $\Omega_{DE}(z)$
in red. } \label{Oede}
\end{figure}

\begin{figure}
\centerline{\includegraphics[width=.9\linewidth,height=.8\linewidth]{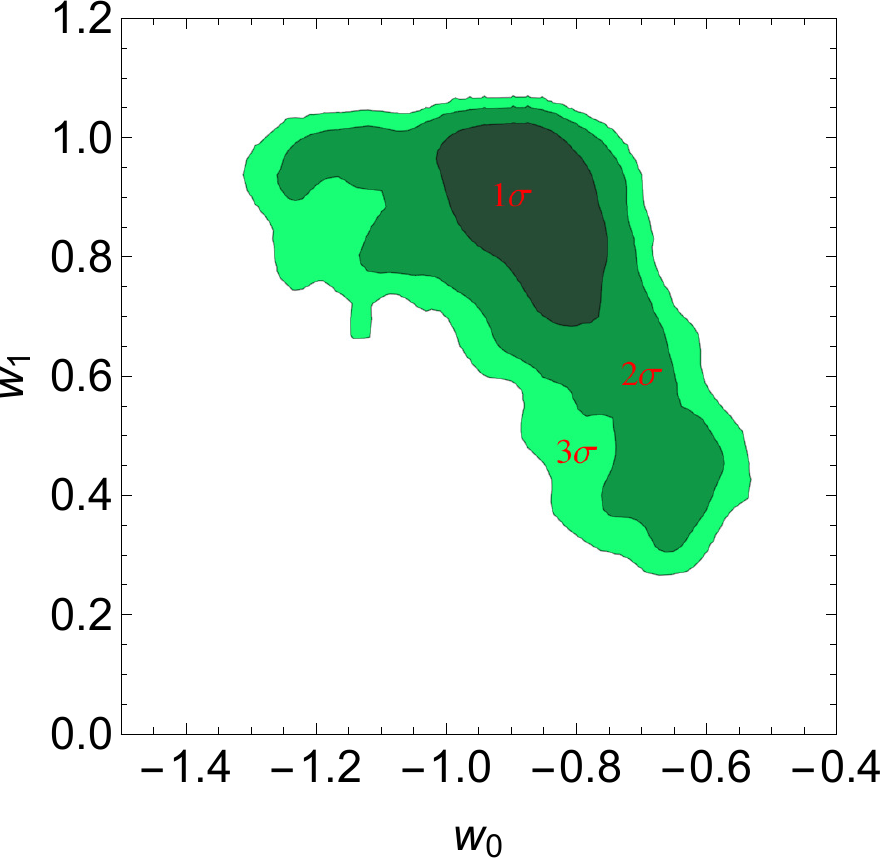}}
\caption{$2D$ confidence
regions in the $w_0$-$w_1$ plane  for the CPL model, obtained
from the GRB Hubble diagram and the $H(z)$ sample. It is worth
noting that the $\Lambda$CDM model of dark energy is disfavoured at more than $3 \sigma$.} \label{cpl_grbhz}
\end{figure}

We sample
the space of parameters by  running five parallel chains and use the Gelman- Rubin diagnostic approach to test the convergence. As a test
probe, it uses the reduction factor $R$, which is the square
root of the ratio of the variance between-chains  and the
variance within-chain.  A large $R$ indicates that the
between-chains variance is substantially greater than the
within-chain variance, so that a longer simulation is needed. We
require that $R$  converges to 1 for each parameter. We set $R -
1$ of order $0.05$, which is more restrictive than the often
used and recommended value $R - 1 < 0.1$ for standard
cosmological investigations.  We discarded the first $30\%$  of
the point iterations at the beginning of any MCMC run, and
thinned the chains that were run many times. We finally
extracted the constrains on cosmographic parameters by coadding
the thinned chains. The histograms of the
parameters from the merged chains were then used to infer median
values and confidence ranges. In Tables
(\ref{tab1}),  (\ref{tabscalar}), and (\ref{tabede}) we present
the results of our analysis. In Fig. (\ref{cpl_grbhz})  we plot a $2D$ confidence region for the CPL
model: it is worth noting that, using only  the GRB Hubble diagram and the
$H(z)$ sample,  the $\Lambda$CDM model of dark
energy is disfavoured at more than $3 \sigma$, as indicated also by the
Hubble diagram of quasars at high redshifts \citep[][]{risaliti1,risaliti2}.
In Figs.  (\ref{wede}) and  (\ref{Oede}) we plot the redshift behaviour of the $\Omega$ parameter and the effective early dark energy (EDE) EOS, corresponding to the best fit values of the parameters, obtained in our statistical analysis.

\begin{table*}
\begin{center}
\resizebox{11cm}{!}{
\begin{tabular}{cccccccccc}
\, & \multicolumn{7}{c}{\bf CPL Dark Energy}   \\
\, & \, & \, & \, & \, & \, & \, & \, & \,   \\
\hline
\, & \, & \, & \, & \, & \, & \, & \, & \,   \\
$Id$ & $\langle x \rangle$ & $\tilde{x}$ & $68\% \ {\rm CL}$  & $95\% \ {\rm CL}$ &  $\langle x \rangle$ & $\tilde{x}$ & $68\% \ {\rm CL}$  & $95\% \ {\rm CL}$ \\
\hline \hline
\, & \, & \, & \, & \, & \, & \, & \, & \,   \\
\hline \, & \multicolumn{4}{c}{SNIa /GRBs/H(z)}  \, & \multicolumn{4}{c}{GRBs/H(z)}
 \\
\hline
\, & \, & \, & \, & \, & \, & \, & \, & \,   \\
$\Omega_m$ &0.29 &0.29& (0.28, 0.30) & (0.27, 0.31) &0. 17&0.18& (0.16, 0.195) & (0.15, 0.26)  \\
\, & \, & \, & \, & \, & \, & \, & \, & \,  \\
$w_0$ &-1.03& -1.02& (-1.1, -0.96) & (-1.14,  -0.88)  &-0.84& -0.858& (-0.94, -0.74) & (-1.07, -0.68) \\
\, & \, & \, & \, & \, & \, & \, & \, & \, &  \\
$w_1$ &0.03&0.03 & (-0.15,0.24) & (-0.32, 0.38) &0.8& 0.86& (0.63, 0.95) & (0.39, 0.99)  \\
\, & \, & \, & \, & \, & \, & \, & \, & \, &  \\
$h$ &0.69& 0.69 & (0.68, 0.70) & (0.67, 0.71)  &0.67& 0.67 & (0.65, 0.7) & (0.62, 0.74)  \\
\, & \, & \, & \, & \, & \, & \, & \, & \, &  \\
\hline
\end{tabular}}
\end{center}
\caption{Constrains on the CPL parameters from different data: combined SNIa  and  GRB Hubble diagrams,  and
$H(z)$ data sets (Left Panel); and GRB Hubble diagram and
$H(z)$ data sets (Right Panel ). Columns show the mean $\langle x \rangle$ and median $\tilde{x}$ values  and the $68\%$ and $95\%$
confidence limits. } \label{tab1}
\end{table*}


\begin{table*}
\begin{center}
\resizebox{15cm}{!}{
\begin{tabular}{cccccccccc}
\, & \multicolumn{7}{c}{\bf Scalar Field Quintessence}   \\
\, & \, & \, & \, & \, & \, & \, & \, & \,   \\
\hline
\, & \, & \, & \, & \, & \, & \, & \, & \,   \\
$Id$ & $\langle x \rangle$ & $\tilde{x}$ & $68\% \ {\rm CL}$  & $95\% \ {\rm CL}$ &  $\langle x \rangle$ & $\tilde{x}$ & $68\% \ {\rm CL}$  & $95\% \ {\rm CL}$ \\
\hline \hline
\, & \, & \, & \, & \, & \, & \, & \, & \,   \\
\hline \, & \multicolumn{4}{c}{SNIa /GRBs/H(z)}  \, & \multicolumn{4}{c}{{\bf GRBs/H(z)}}
 \\
\hline
\, & \, & \, & \, & \, & \, & \, & \, & \,   \\
 $\mathcal{ H}_0$ & 0.98 &0.98& (0.97, 1.0) & (0.95, 1.03) &0. 97&0.98& (0.96, 1.05) & (0.95, 1.1)  \\
\, & \, & \, & \, & \, & \, & \, & \, & \,  \\
$h$ &0.69& 0.69 & (0.68, 0.70) & (0.67, 0.71)  &0.67& 0.67 & (0.65, 0.7) & (0.62, 0.74)  \\
\, & \, & \, & \, & \, & \, & \, & \, & \,   \\
\hline
\end{tabular}}
\end{center}
\caption{ Constrains on the scalar field
parameters from different data: combined SNIa and  GRB Hubble
diagrams,  and $H(z)$ data sets (Left Panel); and GRB Hubble
diagram and $H(z)$ data sets (Right Panel ). Columns show the
mean $\langle x \rangle$ and median $\tilde{x}$ values  and the
$68\%$ and $95\%$ confidence limits. It is worth noting that
$\Omega_m=0.24\pm 0.02$ in the case of the SNIa /GRBs/H(z)
samples, and  $\Omega_m=0.25\pm 0.03$ for the GRBs/H(z)
samples. } \label{tabscalar}
\end{table*}

\begin{table*}
\begin{center}
\resizebox{15cm}{!}{
\begin{tabular}{cccccccccc}
\, & \multicolumn{7}{c}{\bf Early Dark Energy}   \\
\, & \, & \, & \, & \, & \, & \, & \, & \,   \\
\hline
\, & \, & \, & \, & \, & \, & \, & \, & \,   \\
$Id$ & $\langle x \rangle$ & $\tilde{x}$ & $68\% \ {\rm CL}$  & $95\% \ {\rm CL}$ &  $\langle x \rangle$ & $\tilde{x}$ & $68\% \ {\rm CL}$  & $95\% \ {\rm CL}$ \\
\hline \hline
\, & \, & \, & \, & \, & \, & \, & \, & \,   \\
\hline \, & \multicolumn{4}{c}{SNIa /GRBs/H(z)}  \, & \multicolumn{4}{c}{GRBs/H(z)}
 \\
\hline
\, & \, & \, & \, & \, & \, & \, & \, & \,   \\
$\Omega_m$ &0.28 &0.28& (0.26, 0.30) & (0.25, 0.32) &0. 17&0.18& (0.16, 0.195) & (0.15, 0.26)  \\
\, & \, & \, & \, & \, & \, & \, & \, & \,  \\
$w_0$ &-0.62& -0.6& (-0.7,  -0.53) & (-0.95,  -0.5)  &-0.84& -0.858& (-0.94, -0.74) & (-1.07, -0.68) \\
\, & \, & \, & \, & \, & \, & \, & \, & \, &  \\
$\Omega_{e}$ & 0.003&0.003 &  (0.001,0.07) & (0.005, 0.09) &0.008& 0.009& (0.0006, 0.0095) & (0.0004, 0.0015)  \\
\, & \, & \, & \, & \, & \, & \, & \, & \, &  \\
$h$ &0.71& 0.71 & (0.7, 0.72) & (0.69, 0.73)  &0.67& 0.67 & (0.65, 0.7) & (0.62, 0.74)  \\
\, & \, & \, & \, & \, & \, & \, & \, & \, &  \\
\hline
\end{tabular}}
\end{center}
\caption{Constrains on the Early Dark Energy  parameters from different data: combined SNIa  and  GRB Hubble diagrams,  and
$H(z)$ data sets (Left Panel); and GRB Hubble diagram and
$H(z)$ data sets (Right Panel ). Columns show the mean $\langle x \rangle$ and median $\tilde{x}$ values  and the $68\%$ and $95\%$
confidence limits. } \label{tabede}
\end{table*}

\begin{figure}
\centerline{\includegraphics[width=.9\linewidth,height=.8\linewidth]{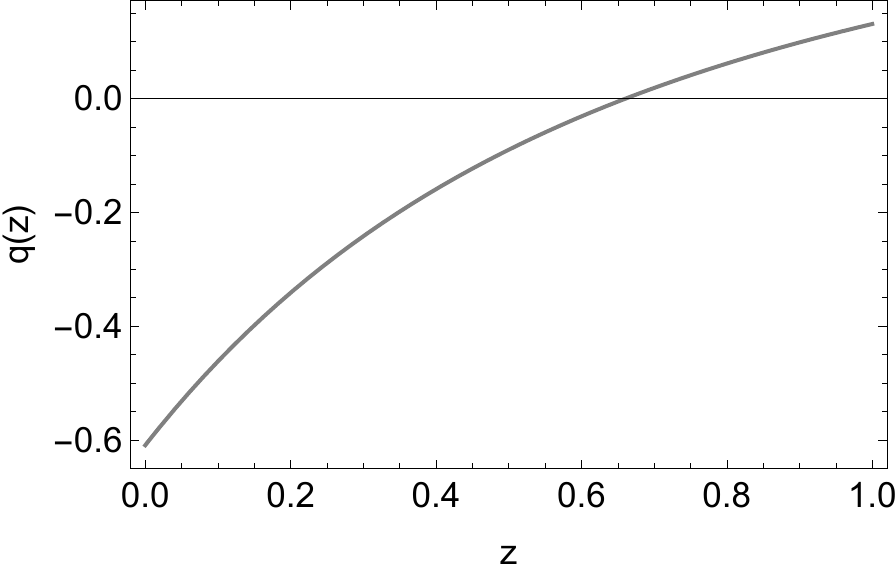}}
\caption{Redshift dependence of the deceleration
parameter $q(z)$ for the CPL model, corresponding to the best
values of the relative parameters obtained combining the GRB
Hubble diagram and the $H(z)$ measurements.} \label{qzcplgrb}
\end{figure}

\begin{figure}
\centerline{\includegraphics[width=.9\linewidth,height=.8\linewidth]{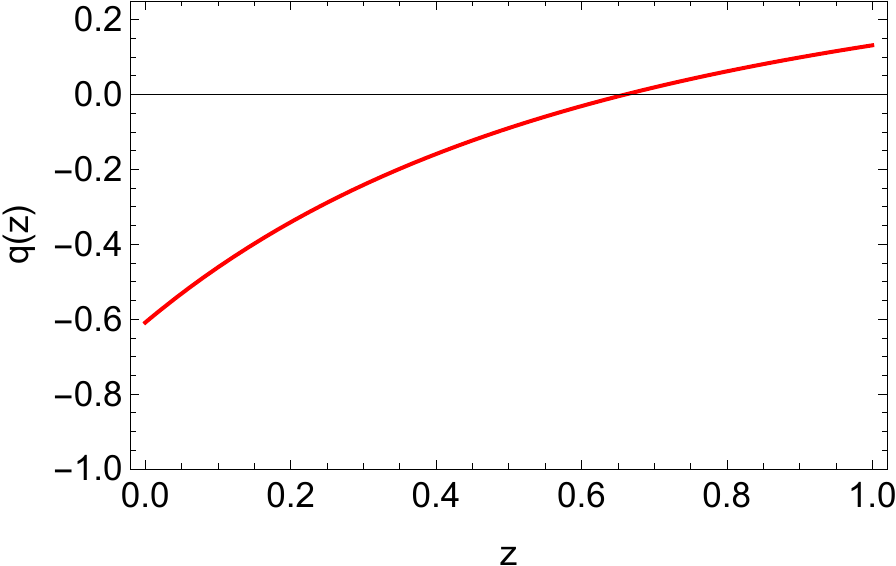}}
\caption{Redshift dependence of the deceleration
parameter $q(z)$ for the scalar field model, corresponding to
the best values of the relative parameters obtained combining
the GRB Hubble diagram and the $H(z)$ measurements.}
\label{qzexpgrb}
\end{figure}

\begin{figure}
\centerline{\includegraphics[width=.9\linewidth,height=.8\linewidth]{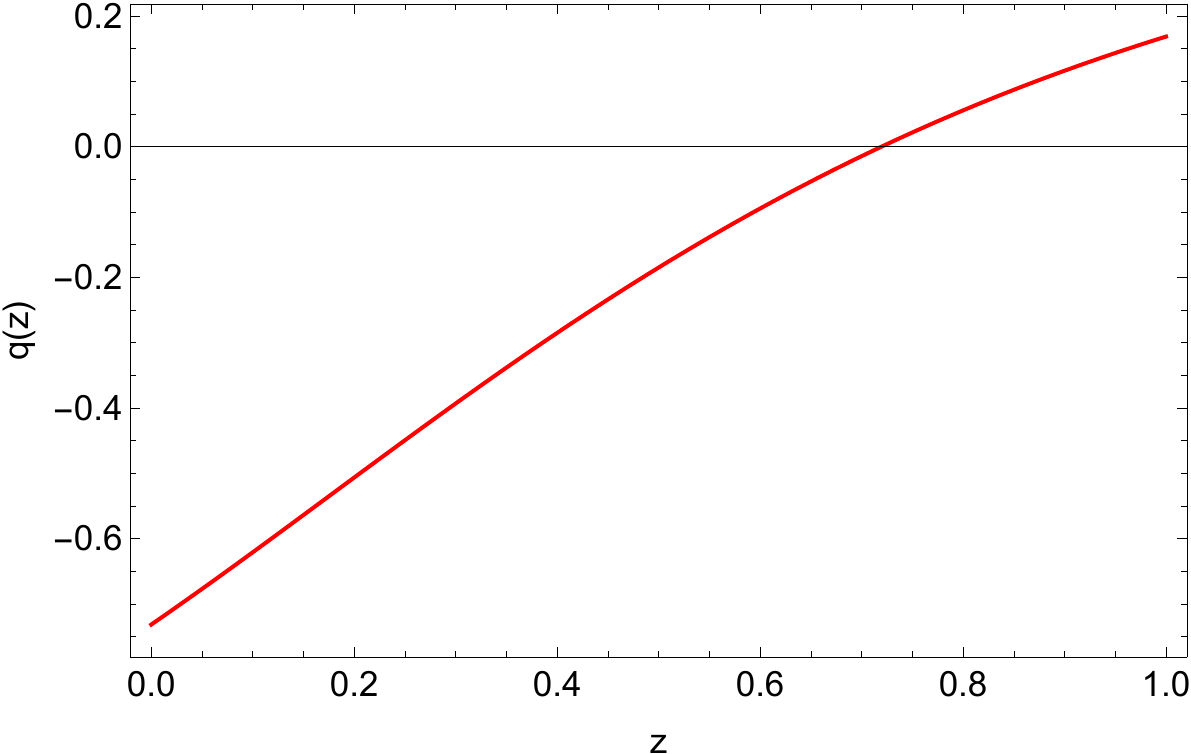}}
\caption{Redshift dependence of the deceleration
parameter $q(z)$ for the EDE model, corresponding to the best
values of the relative parameters obtained combining the GRB
Hubble diagram and the $H(z)$ measurements.} \label{qzedegrb}
\end{figure}

It is well known that the likelihood-statistics
alone does not provide an effective way to compare different
cosmological models. In this section we compare the different
models presented in the previous sections and check if we can
discriminate between them. We use the Akaike Information
Criterion (AIC) \citep[][]{aic}, \citep[][]{aic2}, and its indicator
\begin{equation} \label{aic}
AIC=-2 \ln{\cal{L}_{\bf max} }+  2 n_p +\frac{2 n_p (n_p+1)}{N_{tot}-n_p-1}\,,
\end{equation}
where $N_{tot}$ is the  total number of data and $n_p$ the number of free parameters (of the cosmological model) .
 It turns out that the lower is the value of AIC the better is  the fit to  the data. To compare different cosmological models we introduce the difference
$ \Delta_{AIC} = AIC_{model} - AIC_{min}$. This difference corresponds to different cases:
$4 < \Delta_{AIC} < 7$ indicates a positive evidence against the model with higher value of $AIC_{model}$,
while $\Delta_{AIC} \geq 10 $ indicates a strong evidence. $ \Delta_{AIC} \leq 2$ is an indication that
the two  models are consistent. In our case we have found that the model with the lower AIC is the exponential scalar field:
it turns out that $ \Delta_{AIC} \simeq 5 $ if we consider the CPL model and  $ \Delta_{AIC} = 9 $ for the  early dark energy.
Moreover,  it turns out that also without the
SNIa data, combining the GRB Hubble diagram with the $H(z)$ compilation, it is possible to set the transition region from the decelerated to the accelerated expansion in
all the tested cosmological models, as indicated in Figs. (\ref{qzcplgrb}), (\ref{qzexpgrb}), and  (\ref{qzedegrb}).

\section{Prospects with THESEUS}
\begin{figure}
\centerline{\includegraphics[width=.9\linewidth,height=.8\linewidth]{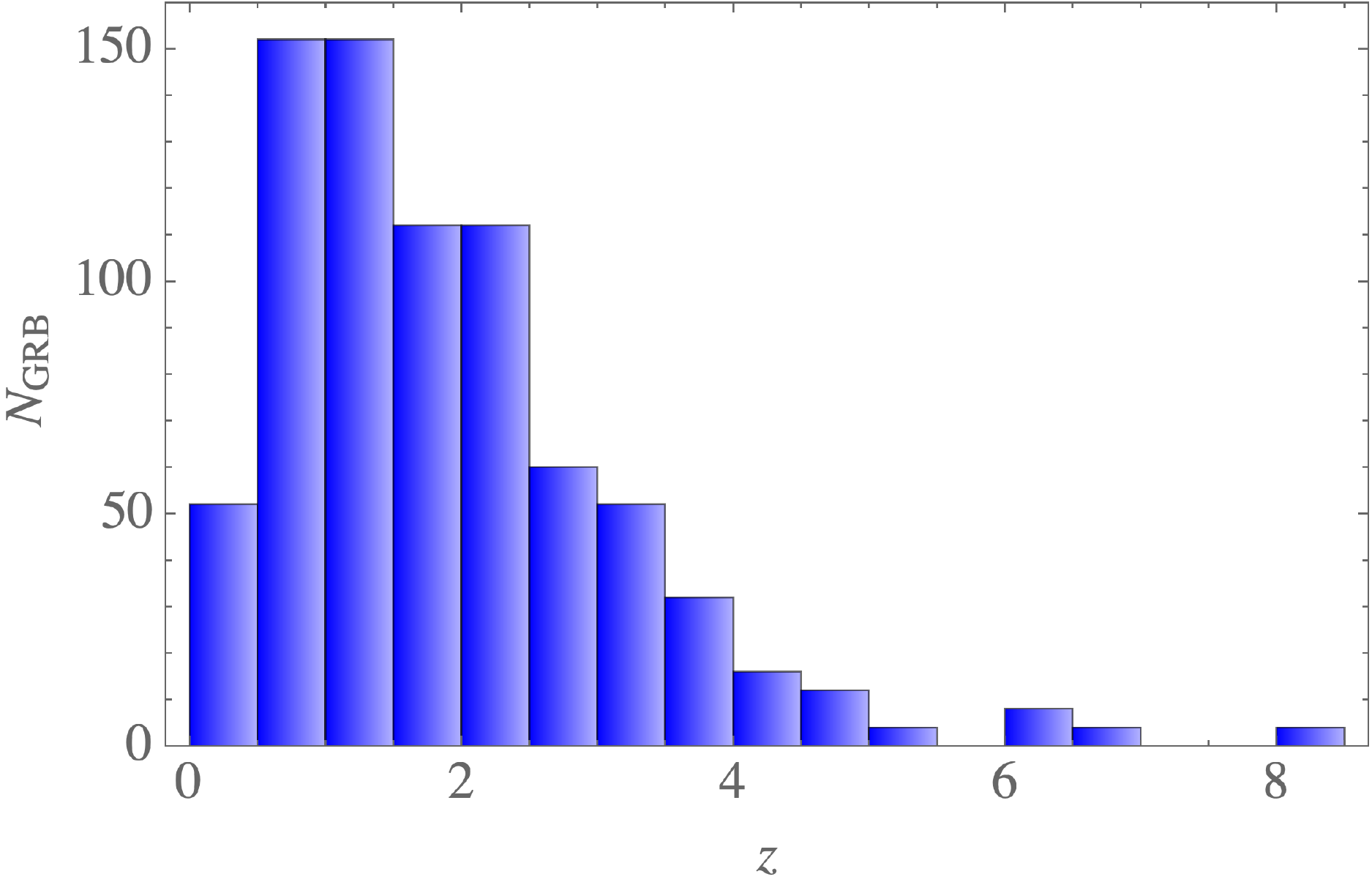}}
\caption{Simulated data for a THESEUS like mission: we plot the GRB redshift distribution of the sample of 772 GRBs used in our analysis.}\label{N-GRBmock}
\end{figure}

So far we have shown that the  $E_{\rm p,i}$ -- $E_{\rm iso}$
correlation has significant implications  for the use of GRBs in
cosmology and therefore GRBs are powerful cosmological probe,
complementary to other probes. GRB missions, like,
the proposed THESEUS observatory \citep[][]{Amati_theseus, cordier18}, will
substantially increase the number of GRBs that could be used to construct
the $E_{\rm p,i}$ -- $E_{\rm iso}$  correlation  up to redshift
$ z \simeq 10$ and will allow a better calibration of this correlation.
Here we consider a simulated sample
of 772 objects to constrain our models, their redshift distribution is shown in Fig. (\ref{N-GRBmock}).

\begin{figure}
\begin{minipage}[b]{0.9\linewidth}
\centerline{\includegraphics[width=9 cm,height=9 cm]{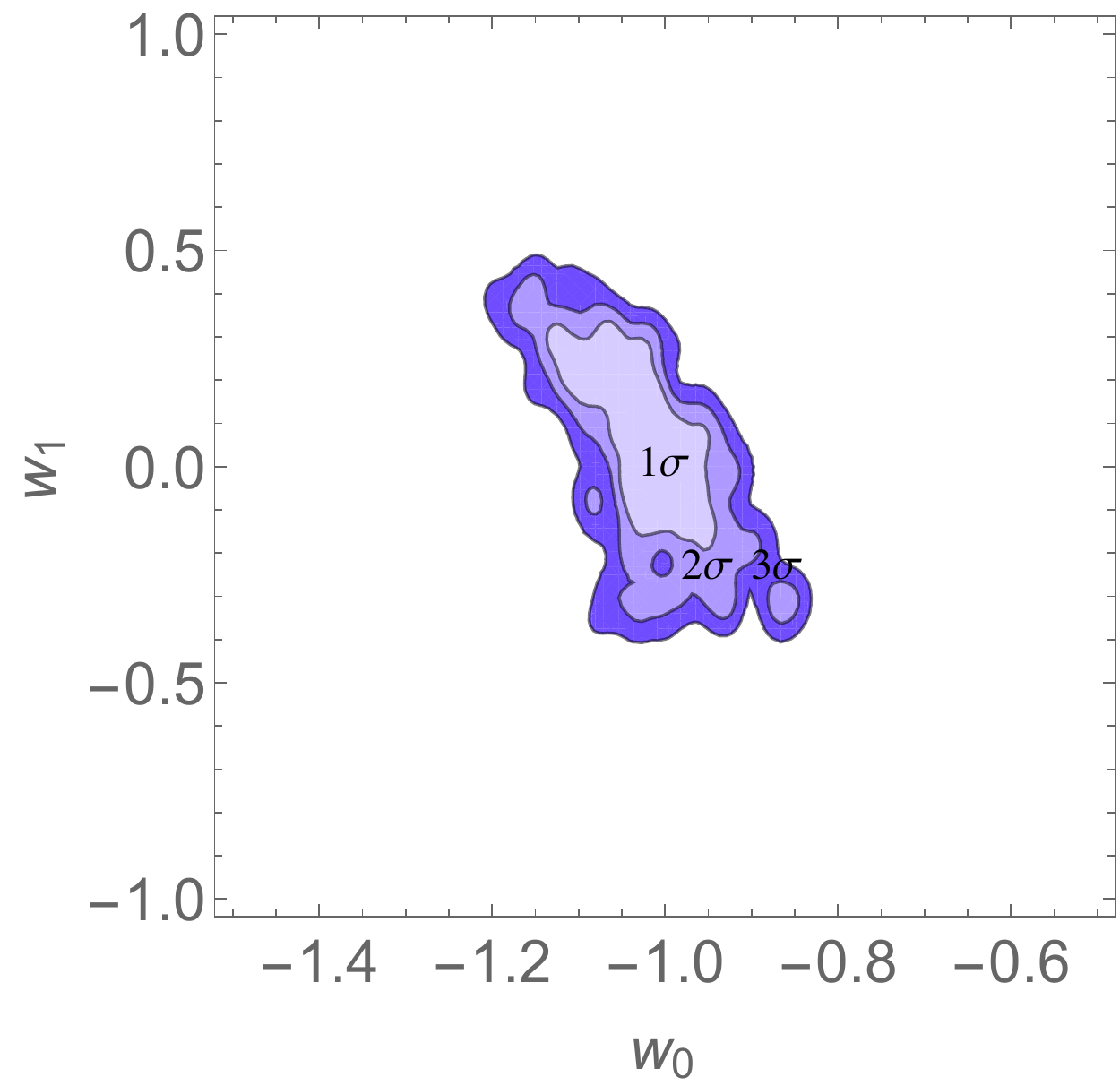}}
 \end{minipage}
 \caption{$2D$ confidence
regions in the $w_0$-$w_1$ plane for the CPL model, obtained
from the  full real datasets.}
\label{2dwow1grbhzsim}
\end{figure}

\begin{figure}
 \begin{minipage}[b]{0.9\linewidth}
\centerline{\includegraphics[width=9 cm,height=9 cm]{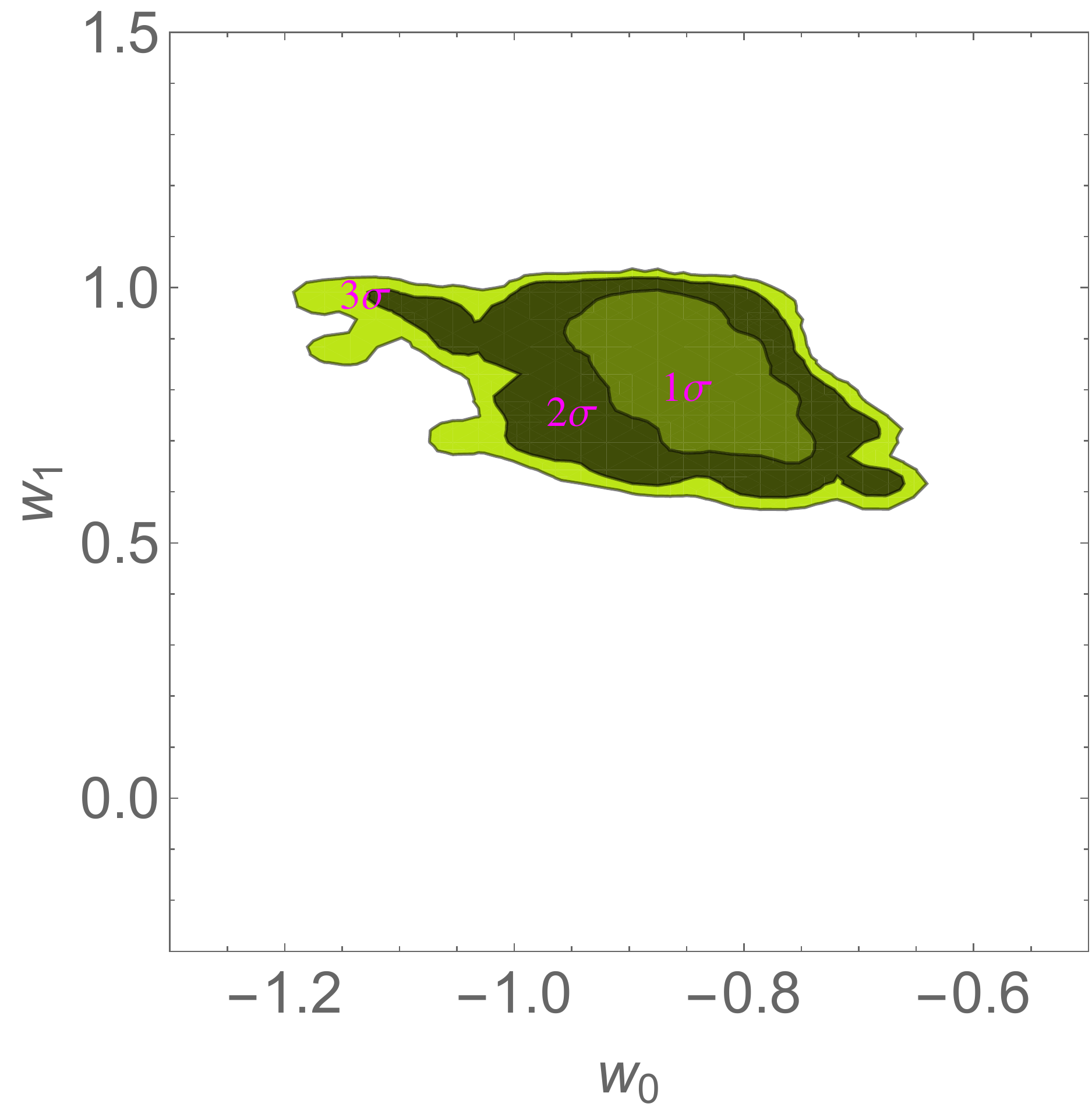}}
 \end{minipage}
\caption{$2D$ confidence
regions in the $w_0$-$w_1$ plane for the CPL model, obtained
from the  simulated GRBs HD and the $H(z)$ measurements (bottom
panel) and the full real datasets (upper panel).}
\label{2dwow1grbhzsimb}
\end{figure}

\begin{figure}
\begin{minipage}[b]{0.9\linewidth}
\centerline{\includegraphics[width=8 cm,height=8 cm]{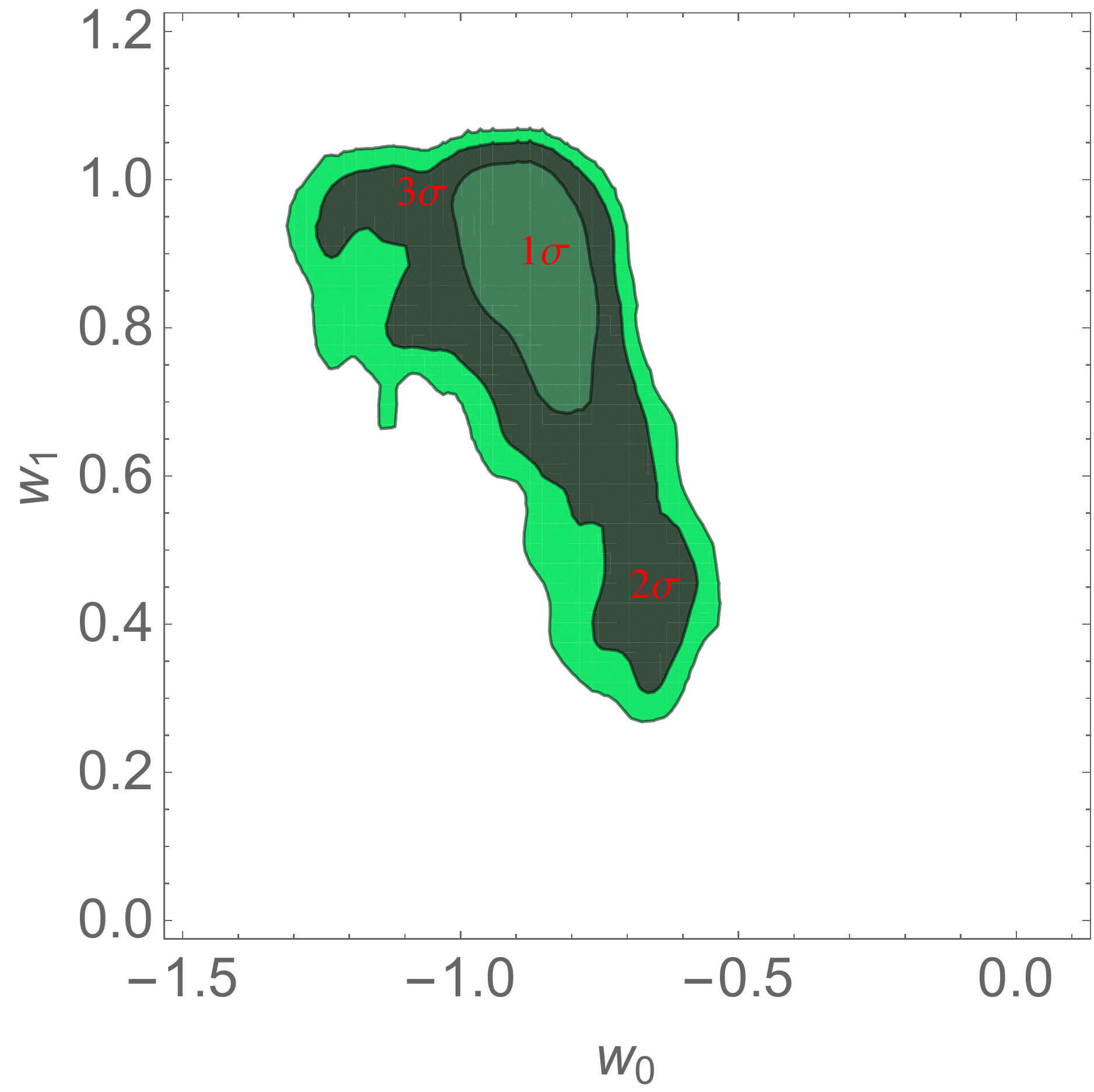}}
 \end{minipage}
\caption{$2D$ confidence regions in the $w_0$-$w_1$ plane for the CPL model,
obtained from the GRBs HD and the $H(z)$ measurements .}
\label{2dwow1grbhzsim2}
\end{figure}

These simulated data
sets have been obtained by implementing the Monte Carlo approach
and  taking into account the slope, normalization, dispersion of
the observed $E_{\rm p,i}$ and $E_{\rm iso}$ correlation, the
distribution of the uncertainties in the measured values
of $E_{\rm p,i}$ and $E_{\rm iso}$, and finally the observed
redshift distribution of GRBs.  In this simulations we took into account the sensitivity limits and spectroscopy thresholds and sensitivity of the THESEUS monitors (SXI and XGIS). This mock sample is based on parameters of the observed sample and corresponds to the actual data sets and
to the data sets expected to be available within 3-4 years from
THESEUS. The cosmological parameters assumed for the simulations are the median, or average, values found in the reported analysis on real data and, indeed, Tables (\ref{tabcplsimul}), and (\ref{tabexpsimul}) show that the analysis on the simulated data recover very well these assumed cosmological parameters.  It turns out that
with our mock sample of  GRBs we are able to constrain much better the cosmological
parameters. Actually, in Figs. (\ref{2dwow1grbhzsim}),  (\ref{2dwow1grbhzsimb}) and
(\ref{2dwow1grbhzsim2}), we show the  $2D$ confidence regions in
the $w_0$-$w_1$ plane  for the CPL model, obtained from the
simulated GRB Hubble diagram and the $H(z)$ sample, compared
with the same confidence region obtained from the real datasets, it
turns out that the evolving dark energy,  described by the exponential scalar field potential is the favoured model. Moreover, in the case of the CPL model, we tested the efficiency of our probes looking at the Figure of Merit (FoM), that is the inverse of area of the $w_0-w_1$ contour: for this purpose we simulate new mock samples of the CPL cosmological model corresponding to the best fit values in Table (\ref{tabcplsimul}), with the  probability density function for the distribution in redshift corresponding to the histograms in Figs. (\ref{N-GRBmock}) and (\ref{N-GRBmock2}), and containing respectively 772 and 1500 objects. To each simulated GRB we estimate the error on the distance modulus as:
\begin{figure}
\centerline{\includegraphics[width=.9\linewidth,height=.8\linewidth]{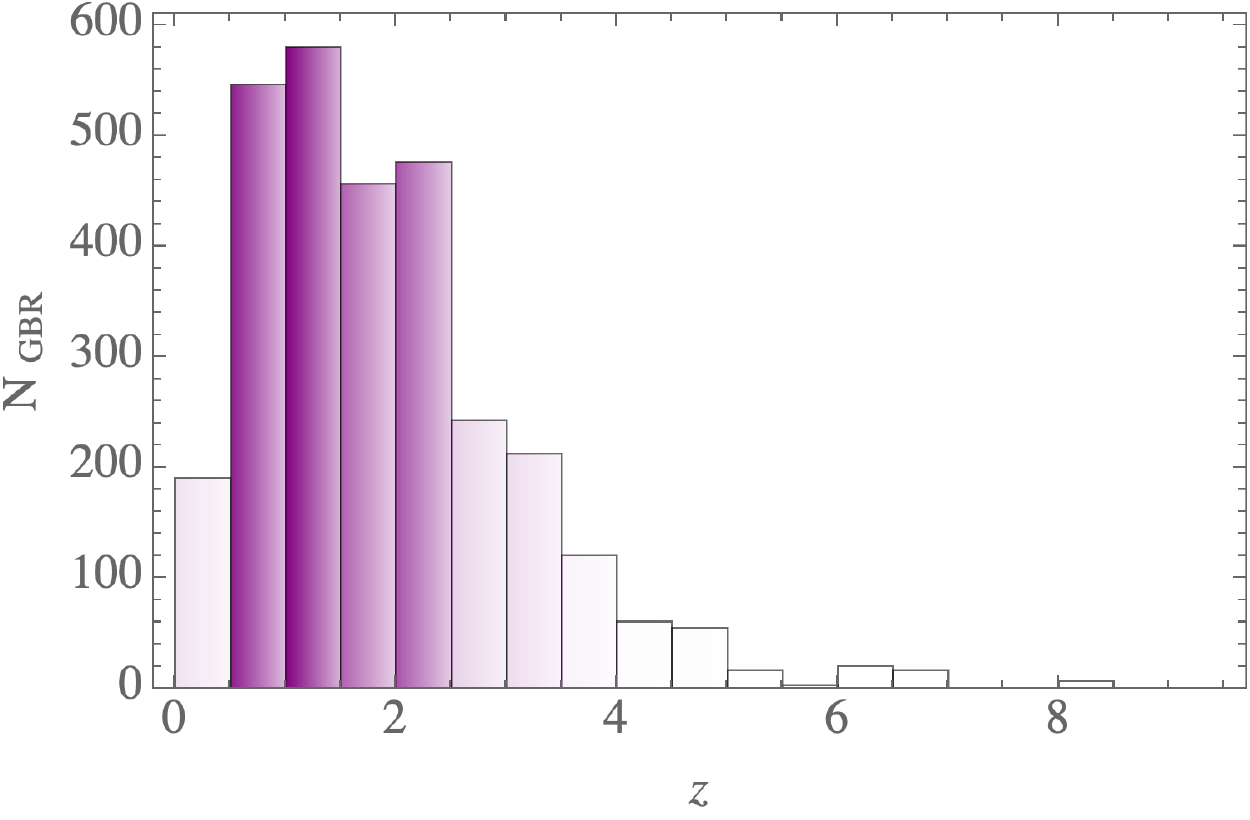}}
\caption{ Simulated data for a THESEUS like mission: we plot the GRB redshift distribution of the sample of 1500 GRBs used in our analysis.}
\label{N-GRBmock2}
\end{figure}
\begin{equation}
\sigma_{\mu}(z) = \sqrt{\sigma^2_{sys}+ \left(\frac{z}{z_{max}}\right)^2 \sigma^2_m}\,.
\label{ersim}
\end{equation}
Here $z_{max }$ is the maximum redshift of the sample, and $ \sigma_{sys}$ the intrinsic scatter. In our case $\left(z_{max }, \sigma_{sys},  \sigma_m\right) =\left(8.5, 0.3, 0.05\right)$.
We get $FoM= 1.9$ when using 772 GRBs, and it increases up to $FoM=4.8$ if 1500 GRBs are used, thus confirming that with future data we will able to better constrain the dark energy EOS as described by the CPL parametrization.

\begin{table*}
\begin{center}
\resizebox{10 cm}{!}{
\begin{tabular}{cccccc}
\,& \multicolumn{4}{c}{\bf CPL Dark Energy}   \\
\, & \, & \, & \, & \, &   \\
\hline
\, & \, & \, & \, & \, &   \\
$Id$ & $\langle x \rangle$ & $\tilde{x}$ & $68\% \ {\rm CL}$  & $95\% \ {\rm CL}$ \\
\hline \hline
\, & \, & \, & \, & \, & \\
\hline & \multicolumn{4}{c}{Simulated GRBs/H(z)}
 \\
\hline
\, & \, & \, & \, & \, &   \\
$\Omega_m$ &0.18&0.19& (0.16, 0.21) & (0.15, 0.27)  \\
\, & \, & \, & \, & \, &  \\
$w_0$ &-0.86& -0.86& (-0.94, -0.79) & (-1.07,  -0.72)   \\
\, & \, & \, & \, & \, & \\
$w_1$ &0.82&0.82 & (0.71,0.93) & (0.62, 0.98)   \\
\, & \, & \, & \, & \, & \\
$h$ &0.67& 0.67 & (0.65, 0.69) & (0.61, 0.74)  \\
\, & \, & \, & \, & \, &  \\
\hline
\end{tabular}}
\end{center}
\caption{Constrains on the CPL parameters from our simulated GRB Hubble diagram and $H(z)$ data.} \label{tabcplsimul}
\end{table*}

\begin{table*}
\begin{center}
\resizebox{10 cm}{!}{
\begin{tabular}{cccccc}
\,& \multicolumn{4}{c}{\bf Scalar Field Quintessence}   \\
\, & \, & \, & \, & \, &   \\
\hline
\, & \, & \, & \, & \, &   \\
$Id$ & $\langle x \rangle$ & $\tilde{x}$ & $68\% \ {\rm CL}$  & $95\% \ {\rm CL}$ \\
\hline \hline
\, & \, & \, & \, & \, & \\
\hline & \multicolumn{4}{c}{Simulated GRBs/H(z)}
 \\
\hline
\, & \, & \, & \, & \, &   \\
$\mathcal{ H}_0$ &0.98&0.981& (0.96, 1.00) & (0.94, 1.02)  \\
\, & \, & \, & \, & \, & \\
$h$ &0.68& 0.68 & (0.68, 0.69) & (0.65, 0.70)  \\
\, & \, & \, & \, & \, &  \\
\hline
\end{tabular}}
\end{center}
\caption{Constrains on the scalar field parameters from our simulated GRB Hubble diagram and $H(z)$ data. It turns out that $\Omega_m= 0.26\pm 0.02$}
\label{tabexpsimul}
\end{table*}

\section{Discussion and conclusions}
The  $E_{\rm p,i}$ -- $E_{\rm iso}$ correlation has significant implications  for the use of GRBs  in cosmology to test different models of dark energy beyond the standard $\Lambda$CDM . Here we
considered an {\it extended} $E_{\rm p,i}$ -- $E_{\rm iso}$ correlation, which takes into account possible
redshift evolution effects (related, for instance to the gravitational lensing along the GRB line of sight), parametrized as power law terms. Using recently
updated data set of 193 high-redshift GRBs, we  applied a local regression technique to calibrate the $E_{\rm p,i}$ -- $E_{\rm iso}$ relation.
The values of the calibration parameters are statistically fully consistent with the  results of our  previous work  \citep[][]{MEC11, MGRB1},
and confirm that for this correlation the evolution effects remain within the intrinsic scatter of the correlation around the best fit values, $\sigma_{int}$. This result has been confirmed also with different calibration techniques, based on an approximation function able to reproduce the luminosity distance in different cosmological models.
Moreover in order to further investigate this question here we follow a different approach: in order to remove possible systematics, we apply some filters on the $E_{\rm p,i}$ -- $E_{\rm iso}$ correlation, selecting a subsample where they are homogeneously distributed in the redshift space and we showed that none of the measured values of $E_{\rm p,i}$ -- $E_{\rm iso}$ are  systematically larger at lower redshifts than at higher redshifts. Again we fit the calibration within this smaller sample, and the results are fully consistent with the results obtained on the whole dataset. Therefore these results justify extension  of this calibration to the whole sample, at least on the basis  of  the current stage of knowledge.
The fitted calibration parameters have been used to construct a high redshift
GRB Hubble diagram, which we adopted as a tool  to constrain different cosmological models: to investigate the prospective for high redshift constrains on dark energy models with the $E_{\rm p,i}$ -- $E_{\rm iso}$ correlation is the prime objective of our analysis. We considered the CPL
parametrization of the EOS, an exponential potential of a self interacting scalar field, and,  finally a model with dark energy at early
times. We compare these different models, by using the Akaike Information Criterion (AIC) and its indicator. In our case we have found that the model with the lower AIC is the exponential scalar field: $ \Delta_{AIC} \simeq 5 $ if we consider the CPL model and  $ \Delta_{AIC} = 9 $ for the  early dark energy. Therefore this model is slightly preferred by the present data. Moreover, even if the cosmological constrains from the currently available GRB Hubble diagram are not so restrictive, future GRB missions, like the proposed THESEUS observatory will increase the number of GRB usable to construct the  $E_{\rm p,i}$ -- $E_{\rm iso}$ correlation up to redshift $z \simeq 10$. We actually considered a mock sample, consisting of 772 objects, obtained taking into account the slope, normalization, dispersion of the observed correlation, the distribution of the uncertainties in the measured values
of $E_{\rm p,i}$ and $E_{\rm iso}$, and finally the observed redshift distribution of GRBs.  In these simulations we took into account the sensitivity limits and spectroscopy thresholds and sensitivity of the THESEUS monitors. The mock sample corresponds to the data sets expected to be available within few years from THESEUS. It turns out that
in this case we are able to constrain much better the cosmological parameters, and the exponential scalar field potential is  confirmed as the favourite model. We finally  tested the efficiency of our probes looking at the Figure of Merit (FoM). In the case of the CPL model: we simulated new mock samples  consisting of 772 and 1500 objects respectively, and  we get $FoM= 1.9$ when using 772 GRBs, while it increases by a factor 2.5 if 1500 GRBs are used.  This confirms, indeed, that with future data we will be able to better constrain the evolution of the dark energy EOS, in a
complementary way to type Ia SN, and in synergy with other independent high redshift cosmological probes, as, for instance quasars \citep[][]{Lusso20, Lusso20b}.

\subsection*{Acknowledgments}
MD is grateful to the INFN for financial support through the Fondi FAI GrIV.
EP acknowledges the support of INFN Sez. di Napoli  (Iniziativa Specifica QGSKY ).
LA  acknowledges support by the Italian Ministry for Education,
University and Research through PRIN MIUR 2009 project on "Gamma ray
bursts: from progenitors to the physics of the prompt emission process." (Prot. 2009 ERC3HT).

\section*{Data Availability}
The data underlying this article will be shared on reasonable request to the corresponding author.




\end{document}